\documentclass[preprint]{revtex4}
\usepackage{graphicx}
\usepackage{bm}

\usepackage{mhchem}
\usepackage{bm}
\usepackage{amsmath}
\usepackage{amsfonts}
\usepackage{amssymb}

\usepackage{breqn}

\usepackage{balance}

\usepackage{times,mathptmx}

\newcommand{\Id}{{\bm 1}}

\usepackage{subfigure}

\newcommand{\vpre}{{U_{\rm w, pre}}}
\newcommand{\vpost}{{U_{\rm w, post}}}
\newcommand{\uw}{{U_{\rm w}}}

\newcommand{\be}{\begin{equation}}
\newcommand{\ee}{\end{equation}}
\newcommand{\bea}{\begin{eqnarray}}
\newcommand{\eea}{\end{eqnarray}}

\begin{document}

\title{Deformation and break-up of viscoelastic droplets in confined shear flow}
\author{A.Gupta $^{1}$, M. Sbragaglia $^{1}$\\
$^{1}$ Department of Physics and INFN, University of ``Tor Vergata'', Via della Ricerca Scientifica 1, 00133 Rome, Italy\\}


\begin{abstract}
The deformation and break-up of Newtonian/viscoelastic droplets are studied in confined shear flow.  Our numerical approach is based on a combination of lattice-Boltzmann models (LBM) and finite difference schemes, the former used to model two immiscible fluids with variable viscous ratio, and the latter used to model the polymer dynamics. The kinetics of the polymers is introduced using constitutive equations for viscoelastic fluids with finitely extensible non-linear elastic dumbbells with Peterlin's closure (FENE-P). We quantify the droplet response by changing the polymer relaxation time $\tau_P$, the maximum extensibility $L$ of the polymers, and the degree of confinement, i.e. the ratio of the droplet diameter to wall separation. In unconfined shear flow, the effects of droplet viscoelasticity on the critical Capillary number $\mbox{Ca}_{\mbox{\tiny{cr}}}$ for break-up are moderate in all cases studied. However, in confined conditions a different behaviour is observed: the critical Capillary number of a viscoelastic droplet increases or decreases, depending on the maximum elongation of the polymers, the latter affecting the extensional viscosity of the polymeric solution. Force balance is monitored in the numerical simulations to validate the physical picture.
\end{abstract}

\maketitle{}

\section{Introduction}\label{sec:intro}

Emulsions play an important role in a huge variety of applications, including foods, cosmetics, chemical and material processing~\cite{Larson}. Deformation, break-up and coalescence of droplets occur during flow, and the control over these processes is imperative to synthesize the desired macroscopic behaviour of the emulsion. The problem is also challenging from the theoretical point of view: it is intrinsically multiscale, as it bridges between the ``microscopic'' dynamics of single constituents (i.e. droplets) and the macroscopic behaviour of the emulsion~\cite{TuckerMoldenaers02}. Most of the times, the synthesis of the emulsion takes place in presence of confinement: this is the case of microfluidic technologies, which are gaining importance as a promising route for the emulsion fabrication~\cite{Christopher07,Seeman12}. Moreover, in real processing conditions, relevant constituents have commonly a viscoelastic -rather than Newtonian- nature. The ``single'' droplet problem has been considered to be the simplest model: in the case of dilute emulsions with negligible droplets interactions, the dynamics of a single droplet indeed provides complete information about the emulsion behaviour. Single droplet deformation and break-up have been extensively studied and reviewed in the literature for the case of Newtonian fluids~\cite{Taylor34,Grace,Stone,Rallison,Fischer}. In the classical problem studied by Taylor~\cite{Taylor34}, a droplet (D) with radius $R$, interfacial tension $\sigma$, and viscosity $\eta_D$ is suspended in another immiscible fluid matrix (M) with viscosity $\eta_M$ under the effect of a shear flow with intensity $\dot{\gamma}$. The various physical quantities are grouped in two dimensionless numbers, the Capillary number $\mbox{Ca}=\dot{\gamma} R \eta_M/\sigma$, giving a dimensionless measure of the balance between viscous and interfacial forces, and the viscous ratio $\lambda=\eta_D/\eta_M$, going from zero for vanishing values of the droplet viscosity (i.e. a bubble) to infinity (i.e. a solid particle). Break-up occurs at a critical Capillary number $\mbox{Ca}_{\mbox{\tiny{cr}}}$, which depends on the viscous ratio $\lambda$~\cite{Grace}. In presence of confinement, a third parameter has to be taken into account: that is the confinement ratio, defined as the ratio between the droplet diameter $2R$ and the wall separation $H$~\cite{ShapiraHaber90}. Confinement suppresses break-up for small viscosity ratios $\lambda < 1$, while promoting it for $\lambda >1$. Confinement can promote break-up even of droplets with a viscosity ratio larger than $4$~\cite{Janssen10,Minale08,Vananroye06}, which cannot be broken in unconfined shear flows~\cite{Grace}. It has also been suggested that the conditions of a uniform and confined shear flow can be exploited to generate quasi monodisperse emulsions by controlled break-up~\cite{Sibillo06,Renardy07}. This is supported by experiments and numerical simulations~\cite{Janssen10} where multiple neckings are observed.\\
Viscoelasticity changes droplet deformation as well as the critical Capillary number for break-up. It is generally accepted that viscoelasticity stabilizes  unconfined droplets against break-up~\cite{Flumerfelt72,Elmendorp,Mighri,Lerdwijitjarud03,Lerdwijitjarud04,AggarwalSarkar07,AggarwalSarkar08,GuidoRev,Verhulst09a,Verhulst09b}. It has also been theoretically predicted that viscoelastic effects show up in the droplet deformation in terms of two dimensionless parameters: the Deborah number, $\mbox{De}=\frac{N_1 R}{2 \sigma}\frac{1}{Ca^2}$, where $N_1$ is the first normal stress difference generated in simple shear flow~\cite{bird87}, and the ratio $N_2/N_1$ between the second and first normal stresses difference~\cite{Greco02}. In spite of its relevance, the current understanding of the combined effect of confinement and viscoelasticity on droplets deformation up to and including break-up is rather limited. Experimental data on the dynamics of confined droplets that contain viscoelastic components are rare~\cite{Cardinaels09,Minale10,Cardinaels10,Cardinaels11,Cardinaelsetal11b,Vananroye11b}. Cardinaels {\it et al.}~\cite{Cardinaels09} investigated droplets under confinement for confinement ratios $0.1<2R/H<0.75$, viscosity ratio equal to $\lambda=0.45$ and $\lambda=1.5$, and Deborah number $\mbox{De}=1$. Matrix viscoelasticity has been found to enhance wall effects and good overall agreement was found by comparing experimental data with predictions from theoretical models~\cite{ShapiraHaber90}. Confined droplet relaxation was studied in Cardinaels {\it et al.}~\cite{Cardinaels10}, revealing a complex non trivial interaction between geometrical confinement and component viscoelasticity. Another recent study by Cardinaels {\it et al.}~\cite{Cardinaels11} also analyzed droplet break-up in systems with either a viscoelastic matrix or a viscoelastic droplet. For a viscoelastic droplet the authors report critical Capillary numbers which are similar to those of a Newtonian droplet, whereas matrix viscoelasticity causes break-up at a much lower Capillary number. Issues related to the capability of viscoelasticity to suppress multiple neckings were also discussed. Complementing experimental results with systematic investigations by varying deformation rates and fluid constitutive parameters is of extreme interest. This is witnessed by the various papers in the literature, addressing the effects of viscoelastic components on droplet deformation and break-up in numerical simulations. Transient behaviour and deformation of a two-dimensional Oldroyd-B droplet in a Newtonian matrix were analyzed by Toose {\it et al.}~\cite{Toose} using a boundary-integral method. Ramaswamy \& Leal~\cite{RamaswamyLeal99a,RamaswamyLeal99b} and Hooper {\it et al.}~\cite{Hooper01} used instead a finite-element method to investigate axisymmetric deformation of viscoelastic droplets using FENE-CR and Oldroyd-B equations~\cite{bird87}. They predicted reduced deformation for a viscoelastic droplet in a viscous matrix and enhanced deformation in the reversed case. Pillapakam \& Singh~\cite{PillapakamSingh04} presented finite-element simulations using an Oldroyd-B model. They report a non-monotonic change in deformation for a viscoelastic droplet in a viscous matrix while the reversed case was seen to increase droplet deformation. Yue {\it et al.}~\cite{Yue04,Yueetal05,Yueetal06a,Yueetal06b,Yueetal08,Yueetal12} performed various numerical calculations  based on a diffuse-interface formulation and the Oldroyd-B constitutive equation for the non-Newtonian phase~\cite{Yueetal05}. Such analysis was then extended by Aggarwal \& Sarkar~\cite{AggarwalSarkar07,AggarwalSarkar08} using a 3D front-tracking finite difference numerical method. In the case of a Newtonian droplet in a viscoelastic matrix they found an increased droplet orientation along the flow direction with respect to the Newtonian case, in agreement with previous theoretical predictions and experimental results~\cite{Greco02,MaffettoneGreco04,Minale04}. Furthermore, Aggarwal \& Sarkar~\cite{AggarwalSarkar07} developed a simple force balance ODE model  which predicts the observed scaling of $\mbox{De}$ as a function of $\mbox{Ca}$. At small Deborah numbers, the critical Capillary number was found to increase proportionally with the degree of viscoelasticity, in line with experimental results~\cite{Lerdwijitjarud03}. Some of the numerical simulations in the literature report a non-monotonic change in the steady-state droplet deformation with increasing Deborah number~\cite{Yueetal05,AggarwalSarkar07,AggarwalSarkar08,Mukherjee09}, whereas other investigations of a viscoelastic droplet in a Newtonian matrix and the reversed situation showed a saturation at high Deborah numbers~\cite{Verhulst09a,Verhulst09b}.\\ 
Here, we present a 3D numerical investigation of deformation and break-up of Newtonian/viscoelastic droplets at small Reynolds numbers. The kinetics of the polymers is introduced using constitutive equations for viscoelastic fluids with finitely extensible non-linear elastic dumbbells with Peterlin's closure (FENE-P)~\cite{bird87},  in which the dumbbells can only be stretched by a finite amount, the latter effect parametrized with a maximum extensional length squared $L^2$, hereafter denoted with {\it finite extensibility parameter}. The model supports a positive first normal stress and a zero second normal stress in steady shear flow. It also supports a thinning effect at large shear, although such effect will not be important in our calculations, all the numerical simulations being performed with fluid pairs with nearly constant shear viscosities. We will discuss the interplay between the degree of confinement and the model parameters of the polymer equation, i.e. the relaxation time $\tau_P$ and the maximum elongation of the polymers $L$, by separately tuning the Deborah number and the elongational viscosity of the polymeric phase~\cite{Lindner03}.  We choose a viscous ratio $\lambda=1$, the reason being that is the most studied in the literature~\cite{Taylor34,Grace,Stone,Rallison,Fischer,Minale08,Renardy07}. It is known from the Newtonian case~\cite{Minale08,Janssen10} that confinement hardly affects the critical Capillary number for such viscous ratio. However, as we will see, the effect of viscoelasticity induces significant changes. Issues related to the presence of multiple neckings will also be investigated with the numerical simulations.\\
The paper is organized as follows: in Sec.~\ref{sec:mathematicalformulation} we will present the necessary mathematical background for the problem studied, showing the relevant equations that we integrate in both the matrix and droplet phase. In Sec.~\ref{sec:dropletdeformation} we will present basic benchmark tests to verify the importance of confinement and viscoelasticity in the numerical algorithm. In particular, we will choose a confined case where viscoelasticity is introduced in the matrix phase, so as to produce a sizeable and measurable effect in the droplet orientation that we can benchmark against known results in the literature~\cite{Minale08,Minale10b}. In Sec.~\ref{sec:dropletbreakup} we specialize to the case of droplet viscoelasticity and present a comprehensive study on the interplay between the degree of confinement and the viscoelastic model parameters, i.e. the relaxation time $\tau_P$ and the maximum elongation of the polymer $L$. In Sec.~\ref{sec:forcebalance} we will complement the results discussed in Sec.~\ref{sec:dropletbreakup} by directly monitoring the force balance which is a consequence of the equations of motion. Conclusions follow in Sec.~\ref{sec:conclusions}. The methodology we use is well detailed in another paper~\cite{SbragagliaGupta} and we briefly summarize it in appendix \ref{appendix}.\\

\section{Problem Statement and Mathematical formulation}\label{sec:mathematicalformulation}

Our numerical approach is based on a  combination of lattice-Boltzmann models (LBM) and finite difference schemes, the former used to model two immiscible fluids with variable viscous ratio, and the latter used to model viscoelasticity using the FENE-P constitutive equations. LBM have already been used to model droplet deformation problems~\cite{Xi99,VanDerSman08,Komrakovaa13,Liuetal12} and also viscoelastic flows~\cite{Onishi2,Onishi1,Malaspinas10}. The novelty we offer from the methodological point of view is the exploration of regimes and situations which have not been explored so far in the literature. We focus mainly on the droplet deformation and break-up problems, being the quantitative benchmarks against known analytical results for the rheology of dilute suspensions~\cite{bird87,Herrchen97} present in another dedicated methodological publication~\cite{SbragagliaGupta}. LBM have already been used to model the droplet deformation problems. Three-dimensional numerical simulations of the classical Taylor experiment on droplet deformation~\cite{Taylor34} in a simple shear flow have been performed by Xi \& Duncan~\cite{Xi99} using the so called ``Shan-Chen" approach~\cite{SC93}.  The single droplet problem was also investigated by Van der Sman \& Van der Graaf~\cite{VanDerSman08} using a ``free energy'' LBM. LBM modelling of two phase flows is intrinsically a diffuse interface method and involves a finite thickness of the interface between the two liquids and related free energy model parameters. These numerical degrees of freedom are characterized by two dimensionless numbers, the P\'{e}clet ($\mbox{Pe}$) and Cahn ($\mbox{Ch}$) numbers: the Cahn number is the interface thickness normalized by the droplet radius, whereas the Peclet number is the ratio between the convective time scale and the time scale associated with the interface diffusion. Those parameters have to be chosen within certain ranges to reproduce the correct physical behavior~\cite{VanDerSman08,Komrakovaa13} (see also Appendix \ref{appendix}).  The set-up for the study of break-up is shown in Fig.~\ref{fig:1}. In the droplet phase we integrate both the NS (Navier-Stokes) for the velocity ${\bm u}$ and FENE-P reference equations: 
\begin{eqnarray}\label{EQ}
\rho \left[ \partial_t \bm u + ({\bm u} \cdot {\bm \nabla}) \bm u \right] 
&=&  - {\bm \nabla}P + {\bm \nabla} \cdot \left(\eta_{A} ({\bm \nabla} {\bm u}+({\bm \nabla} {\bm u})^{T})\right)+
              \frac{\eta_P}{\tau_P}{\bm \nabla} \cdot [f(r_P){\bm {\bm {\mathcal C}}}];  
                                                 \label{NS}\\
\partial_t {\bm {\mathcal C}} + (\bm u \cdot {\bm \nabla}) {\bm {\mathcal C}}
&=& {\bm {\mathcal C}} \cdot ({\bm \nabla} {\bm u}) + 
                {({\bm \nabla} {\bm u})^T} \cdot {\bm {\mathcal C}} - 
                \frac{{f(r_P){\bm {\mathcal C}} }- {\Id}}{\tau_P}.
                                                   \label{FENE}
\end{eqnarray}
Here, $\eta_A$ is the dynamic viscosity of the fluid, $\eta_P$ the viscosity parameter for the FENE-P solute, $\tau_P$ the polymer relaxation time, $\rho$ the solvent density, $P$ the solvent pressure, $({\bm \nabla} {\bm u})^T$ the transpose of $({\bm \nabla} {\bm u})$. ${\bm {\mathcal C}} \equiv \langle \mathcal R_i \mathcal R_j\rangle$ is the polymer-conformation tensor, i.e. the ensemble average of the tensor product of the end-to-end distance vector $\mathcal R_i$, which equals the identity  tensor (${\bm {\mathcal C}}=\Id$) at equilibrium. Finally, $f(r_P)\equiv{(L^2 -3)/(L^2 - r_P^2)}$ is the FENE-P potential that ensures finite extensibility, whereas $ r_P \equiv \sqrt{Tr({\bm{\mathcal C}})}$ and $L$ are the length and the maximum possible extension of the polymers~\cite{bird87}, respectively. In the outer matrix phase (indicated with a prime), we consider the equations
\begin{eqnarray}\label{EQB}
\rho^{\prime} \left[ \partial_t \bm u^{\prime} + ({\bm u}^{\prime} \cdot {\bm \nabla}) \bm u^{\prime} \right] 
&=&  - {\bm \nabla}P^{\prime}+ {\bm \nabla} \cdot \left(\eta_{B} ({\bm \nabla} {\bm u}^{\prime}+({\bm \nabla} {\bm u}^{\prime})^{T})\right)+
              \frac{\eta^{\prime}_P}{\tau^{\prime}_P}{\bm \nabla} \cdot [f(r^{\prime}_P){\bm {\bm {\mathcal C}^{\prime}}}];  
                                                 \label{NSb}\\
\partial_t {\bm {\mathcal C}}^{\prime} + (\bm u^{\prime} \cdot {\bm \nabla}) {\bm {\mathcal C}^{\prime}}
&=& {\bm {\mathcal C}}^{\prime} \cdot ({\bm \nabla} {\bm u}^{\prime}) + 
                {({\bm \nabla} {\bm u}^{\prime})^{T}} \cdot {\bm {\mathcal C}^{\prime}} - 
                \frac{{f(r^{\prime}_P){\bm {\mathcal C}^{\prime}} }- {\Id}}{\tau^{\prime}_P}.
                                                   \label{FENEb}
\end{eqnarray} 
with $\eta_{B}$ the matrix shear viscosity. In all the cases, the Navier-Stokes equations are obtained from a lattice Boltzmann model~\cite{Onishi2,Xi99} and immiscibility between the droplet phase and the matrix phase is introduced using the so-called ``Shan-Chen'' model~\cite{SC93,CHEM09}. The methodology is well detailed in another paper~\cite{SbragagliaGupta} and we briefly recall it in appendix \ref{appendix}. In all the numerical simulations presented in this paper, we work with unitary viscous ratio, defined in terms of the total (fluid+polymer) shear viscosity. In particular, when presenting some benchmark studies for droplet deformation (Sec.~\ref{sec:dropletdeformation}), we will choose a case with matrix viscoelasticity ($\eta_P=0$ in Eq.~\eqref{NS}) with $\lambda=\eta_D/\eta_M=\eta_A/(\eta_B+\eta^{\prime}_P)=1$ and polymer concentration $\eta^{\prime}_P/\eta_M \approx 0.4$; all the results for droplet break-up (Sec.~\ref{sec:dropletbreakup}), instead, refer to a case with droplet viscoelasticity ($\eta^{\prime}_P=0$ in Eq.~\eqref{NSb}) with $\lambda=\eta_D/\eta_M=(\eta_A+\eta_P)/\eta_B=1$ and polymer concentration $\eta_P/\eta_D \approx 0.4$. The degree of viscoelasticity is computed from the Deborah number (see also Sec.~\ref{sec:intro}) 
\be\label{De:REAL}
\mbox{De}=\frac{N_1 R}{2 \sigma}\frac{1}{\mbox{Ca}^2}
\ee 
where $\mbox{Ca}$ is always computed in the matrix phase while the first normal stress difference $N_1$ is computed either in the droplet phase (Sec.~\ref{sec:dropletbreakup}) or in the matrix phase (Sec.~\ref{sec:dropletdeformation}), dependently on the case studied. Solving the constitutive equation for steady shear flow, the first normal stress difference $N_1$ for the FENE-P model~\cite{bird87,Lindner03} follows (primed variables replace non-primed variables for matrix phases)
\be\label{N1}
N_1(\tau_P \dot{\gamma},L^2)=8 \frac{\eta_P}{\tau_P} \left(\frac{L^2}{6} \right) \sinh^2 \left(\frac{1}{3} \mbox{arcsinh} \left(\frac{\tau_P \dot{\gamma} L^2}{4} \left(\frac{L^2}{6}\right)^{-3/2}\right) \right).
\ee
In the Oldroyd-B limit ($L^2 \gg 1$) we can use the asymptotic expansion of the hyperbolic functions and we get  $N_1=2 \eta_P \dot{\gamma}^2 \tau_P$ so that 
\be\label{De:OLDROYD}
\mbox{De}=\frac{\tau_P}{\tau_{\mbox{\tiny{em}}}} \frac{\eta_P}{\eta_{M}}
\ee
showing that $\mbox{De}$ is clearly dependent on the ratio between the polymer relaxation time $\tau_P$ and the emulsion time 
\be\label{emulsiontime}
\tau_{\mbox{\tiny{em}}}=\frac{R \eta_{M}}{\sigma}. 
\ee
In the following sections, we report the Deborah number based on the definition \eqref{De:OLDROYD}, as we estimated the difference between~\eqref{De:OLDROYD} and~\eqref{De:REAL} to be at maximum of a few percent for the values of $L^2$ considered.

\section{Steady-State Droplet Deformation/Orientation: Importance of Confinement and Viscoelasticity}\label{sec:dropletdeformation}

\begin{figure}[pth!]
\includegraphics[scale=0.4]{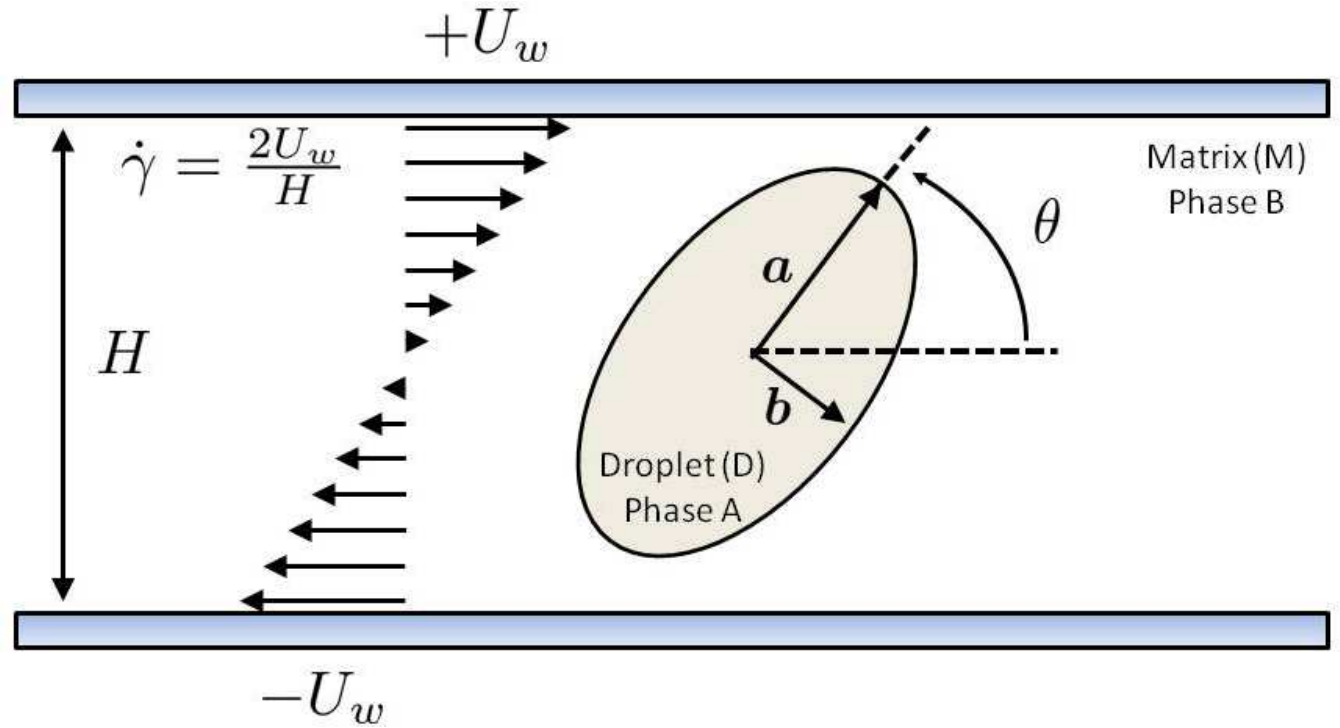}
\caption{Shear plane ($xz$ plane at $y=L_y/2$) view of the numerical set-up for the study of deformation and break-up of confined droplets. A Newtonian droplet (D) (phase $A$) with radius $R$ and shear viscosity $\eta_A$ is placed in between two parallel plates at distance $H$ in a Newtonian matrix (M) (phase $B$) with shear viscosity $\eta_B$. We then add a polymer phase with shear viscosity $\eta_P/\eta_P^{\prime}$ in the droplet/matrix (D/M) phase. We work with unitary viscous ratio, defined in terms of the total (fluid+polymer) shear viscosity: $\lambda=(\eta_A+\eta_P)/\eta_B=1$ in case of droplet viscoelasticity; $\lambda=\eta_A/(\eta_B+\eta^{\prime}_P)=1$ in case of matrix viscoelasticity (see Eqs. (\ref{NS})-(\ref{FENEb})). A shear $\dot{\gamma}=2 \uw/H$ is applied by moving the two plates in opposite directions with velocities $\pm \uw$. The corresponding Capillary number is given in terms of the matrix viscosity and surface tension $\sigma$ at the interface, $\mbox{Ca}=\dot{\gamma} R \eta_M/\sigma$. In order to quantify the deformation of the droplet, we study the deformation parameter $D=(a-b)/(a+b)$, where $a$ and $b$ are the droplet semi-axes in the shear plane, and the orientation angle $\theta$ between the major semi-axis and the flow direction. Droplet deformation is benchmarked in a case of matrix viscoelasticity in Fig.~\ref{fig:2}. For large $\mbox{Ca}$ the droplet deformation is increased and the droplet breaks at a critical Capillary number $\mbox{Ca}_{\mbox{\tiny{cr}}}$. Droplet break-up will be analyzed for the case of droplet viscoelasticity. \label{fig:1}}
\end{figure}

\begin{figure}[pth!]
\subfigure[\,\,]
{
\includegraphics[scale=0.8]{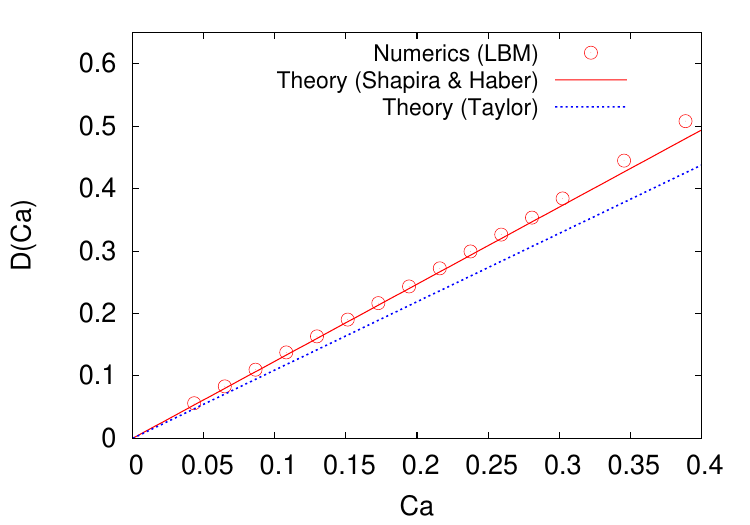}
}
\subfigure[\,\,]
{
\includegraphics[scale=0.8]{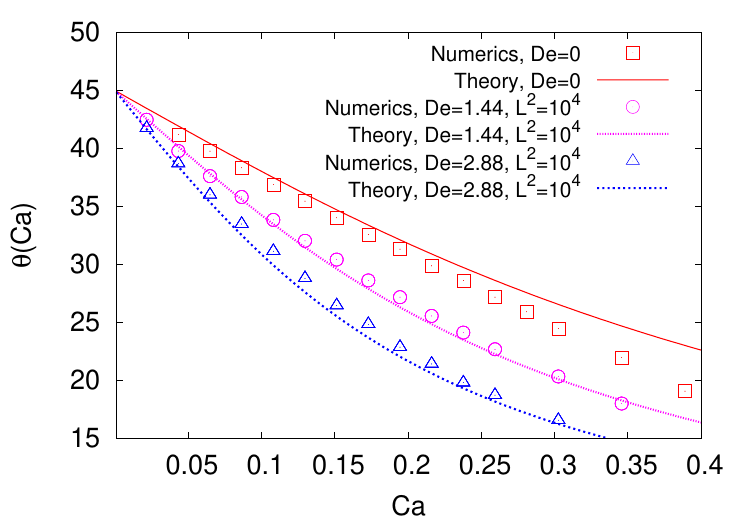}
}
\caption{Panel (a): We report the steady-state deformation parameter $D$ for a Newtonian droplet under steady shear flow as a function of the associated Capillary number $\mbox{Ca}$. The viscous ratio between the droplet phase and the matrix phase is kept fixed to $\lambda=1$. For small $\mbox{Ca}$ the linearity of the deformation is captured by the numerical simulations, but the numerical results overestimate Taylor's prediction (unconfined droplet), being well approximated by the theoretical prediction of Shapira \& Haber for a confined droplet~\cite{ShapiraHaber90}. Panel (b): steady-state orientation angle for a Newtonian droplet immersed in a viscoelastic matrix with Deborah numbers $\mbox{De}=1.44$ and $\mbox{De}=2.88$ and finite extensibility parameter $L^2=10^4$. The results for the corresponding Newtonian system ($\mbox{De}=0$) with the same viscous ratio are also reported. The reference theory comes from the prediction of ``ellipsoidal'' models~\cite{Minale08,Minale10b}, describing the dynamics of a single Newtonian droplet immersed in a viscoelastic matrix, based on the assumption that the droplet deforms into  an ellipsoid.  \label{fig:2}}
\end{figure}

In this section we present benchmark tests of the numerical simulations with regard to the problem of steady-state droplet deformation and orientation in shear flow. In particular, we will show that both the effects of confinement and viscoelasticity are fairly reproduced by our approach. In order to quantify the deformation of the droplet, we study the deformation parameter $D\equiv(a-b)/(a+b)$, where $a$ and $b$ are the droplet semi-axes in the shear plane, and an orientation angle $\theta$ between the major semi-axis and the flow direction (see Fig.~\ref{fig:1}). Taylor's result, based on a small deformation perturbation procedure to first-order, relates the deformation parameter to the Capillary number, 
\be
D = \frac{(19\lambda+ 16)}{(16 \lambda+16)} \mbox{Ca}
\ee
whereas the orientation angle is constant and equal to $\theta=\pi/4$ to first order. Taylor's analysis was later extended by working out the perturbation procedure to second order in $\mbox{Ca}$, which leaves unchanged the expression of the deformation parameter and gives the ${\cal O}(Ca)$ correction to the orientation angle~\cite{Rallison2,Chaffey}. The effects of confinement have been theoretically addressed at first-order by Shapira \& Haber~\cite{ShapiraHaber90,Sibillo06}. They found that the deformation parameter in the confined geometry can be obtained by the unconfined flow expression through a correction in the third power of the ratio between droplet radius at rest $R$ and the gap between the walls $H$
\be\label{deformationSH}
D=\frac{(19\lambda+ 16)}{(16 \lambda+16)}\left[1+C_{\mbox{\tiny{sh}}} \frac{2.5 \lambda+1}{\lambda+1} \left(\frac{R}{H}\right)^3 \right] \mbox{Ca}
\ee
where $C_{\mbox{\tiny{sh}}}$ is a tabulated numerical factor depending on the relative distance between the droplet center and the wall (the value of $C_{\mbox{\tiny{sh}}}$ for droplets placed halfway between the plates is $C_{\mbox{\tiny{sh}}}=5.6996$). Numerical simulations results are presented in Panel (a) of Fig.~\ref{fig:2}. To the best of the authors knowledge, this is the first time that LBM simulations are quantitatively compared with the theoretical prediction by Shapira \& Haber~\cite{ShapiraHaber90,Sibillo06}. In particular, we report the steady-state droplet deformation for a confined shear flow at a given degree of confinement $2R/H = 0.465$ at changing $\mbox{Ca}$.  The droplet radius is $R=30$ lattice cells and the computational domain is  $L_{x} \times L_y \times H = 128 \times 128 \times 128$ lattice cells. The viscous  ratio is $\lambda=1$, the dynamic viscosities in equations (\ref{EQ})-(\ref{EQB})  are $\eta_A=\eta_B=1.75$ lbu (LBM units), and the surface tension  at the non ideal interface is $\sigma=0.1$ lbu. The Capillary number is changed by imposing different velocities at the upper and  lower walls. As we can see, the linearity of the deformation is captured at small  $\mbox{Ca}$, but the numerical results overestimate Taylor's prediction, being well  approximated by the theoretical prediction of Shapira \& Haber for a confined  droplet~\cite{ShapiraHaber90}. As a consequence of this increased droplet deformation  at reduced gap size, elongated shapes are observed at steady-state in confined shear  flow, which would be unstable in the unconfined case~\cite{Sibillo06}.

We next go on by proposing a benchmark test for the viscoelastic effects on shear-induced droplet orientation at  small $\mbox{Ca}$. We prefer to look at the orientation angle $\theta$ (see Fig.~\ref{fig:1}) because  non-Newtonian effects on the steady-state deformation show up at the second order in $\mbox{Ca}$, while the orientation angle has a correction at first order in  $\mbox{Ca}$ ~\cite{Greco02}. Also, we choose to use only matrix viscoelasticity ($\eta_P=0$ in Eq.~\eqref{NS}) because it is known that droplet viscoelasticity has hardly any effect on the steady-state droplet deformation and orientation at small Capillary numbers~\cite{Verhulst09a,Verhulst09b}.  As a reference theory, to test both confinement and viscoelastic effects,  we refer to the model proposed by Minale, Caserta \& Guido~\cite{Minale10}. This model belongs to the family of ``ellipsoidal'' models~\cite{Minale10b}, which were originally introduced to describe the dynamics of a single Newtonian droplet immersed in a Newtonian matrix subjected to a generic flow field, based on the assumption that the droplets deform into an ellipsoid. The steady-state predictions of such models for small Capillary numbers are  constructed in such a way to recover the exact perturbative result, i.e. Taylor's result for an unconfined droplet~\cite{MaffettoneMinale98} or the Shapira \& Haber result for a confined droplet~\cite{Minale08}. Recently, extension of ellipsoidal  models have been proposed also for non-Newtonian fluids. In particular, Minale ~\cite{Minale04} proposed a model which recovers the small deformation steady-state theory developed by Greco~\cite{Greco02} to predict the deformation of a droplet made of a second-order fluid. Minale, Caserta \& Guido~\cite{Minale10} recently generalized the work by Minale~\cite{Minale04,Minale08} to study the effects of confinement in non-Newtonian systems. With respect to the Newtonian case studied in Panel (a) of Fig. \ref{fig:2}, we leave all the parameters unchanged, with the only  difference that we switch on the polymeric viscosity $\eta^{\prime}_P=0.69333$ lbu and lower the solvent matrix viscosity $\eta_{B}$ so as to leave the total viscous ratio $\lambda=\eta_D/\eta_M=\eta_A/(\eta_B+\eta^{\prime}_P)=1$ unchanged. In Panel (b) of Fig.~\ref{fig:2} we report the steady-state orientation angle for a Newtonian droplet immersed in a non-Newtonian matrix with $\mbox{De}=1.44, 2.88$ and $L^2=10^4$. The results for the corresponding Newtonian system ($\mbox{De}=0$)  with the same viscous ratio are also reported. The value of $L^2$ is chosen to avoid  thinning effects in the viscoelastic behaviour which would complicate the quantitative  matching between the reference theory~\cite{Minale10} and the simulations. The effect of viscoelasticity is clearly visible: if compared with the Newtonian case ($\mbox{De}=0$), viscoelasticity promotes stronger alignment in the flow direction and the numerical results are well in agreement with the ellipsoidal model by Minale, Caserta \& Guido~\cite{Minale10} for all the Deborah numbers considered.

\section{Effects of Droplet Viscoelasticity on Critical Capillary Number}\label{sec:dropletbreakup}

In this section we report the results for the critical Capillary number for various confinement  ratios and Deborah numbers. We will be mainly interested in droplet viscoelasticity, which is obtained by setting $\eta^{\prime}_P=0$ in eq.~\eqref{FENEb}. A  complementary study regarding the role of matrix viscoelasticity will be published in a future paper. In all the cases discussed in this section, a spherical droplet is initially placed halfway  between the walls. The critical Capillary number is computed by identifying the pre-critical  ($\vpre$) and the post-critical wall velocity ($\vpost$), i.e. the largest (smallest) wall velocity for which the droplet is stable (breaks). All the simulations described refer to the cases with polymeric relaxation times ranging in the interval $0 \le \tau_{P} \le 7000$ lbu and finite extensibility parameter $10^2 \le L^2 \le  10^4$, corresponding to Deborah numbers ranging in the interval $0 \le De \le 2$. The numerical simulations have been carried out in three dimensional domains  $L_{x} \times L_y \times H$. The droplet radius $R$ and the vertical gap $H$ have been changed in the ranges $50 \le R \le 60 $ lattice cells and $128 \le H \le 256 $ lattice cells to achieve different confinement ratios $2R/H$. The stream-flow length $L_x$ is varying in the range $1024 \le L_{x} \le 1356$ lattice cells, depending on the droplet elongation properties, while the transverse-flow length $L_y$ is resolved with $128$ lattice cells. Periodic conditions are applied in the stream-flow and in the transverse-flow directions. The droplet is subjected to a linear shear flow $u_x=\dot{\gamma} z$, $u_y=u_z=0$, with the shear introduced with two opposite velocities in the stream-flow direction ($-u_x(x,y,z=0)=+u_x(x,y,z=H)=\uw$) at the upper ($z=H$) and lower wall ($z=0$).  The main simulation parameters are summarized in table \ref{table:para}.


\begin{table}\label{table}
   \begin{tabular}{@{\extracolsep{\fill}} |c|c|c|c|c|c|c|c|c|c|c|c|c|}
    \hline
    $2R/H$ & $L_{x} \times L_y \times H$ & $R$ &  $\eta_A$ & $\eta_B$  & $\eta_P$ & $\tau_P$ & $De$ & $L^2$ & $\vpre$ & $\vpost$ \\
   & cells & lbu & lbu & lbu & lbu & lbu & & & lbu & lbu \\
   \hline \hline
    $0.4$ & $1024 \times 128 \times 256$ & $50$ & $1.75$ & $1.75$ & $0.00$ &  $   $	&	$   $ & $ $ & $0.04$ & $0.0425$ \\
    $0.4$ & $1024 \times 128 \times 256$ & $50$ & $1.05$ & $1.75$ & $0.69$ &  $5-50 \times 10^2$ & $0.2-2.0$	&	$10^2$ & $0.04-0.0425$ & $0.0425-0.045$ \\
    $0.4$ & $1024 \times 128 \times 256$ & $50$ & $1.05$ & $1.75$ & $0.69$ &  $50 \times 10^2$	& $2.0$ &	$10^4$ & $0.0425$ & $0.045$ \\
\hline
    $0.45$ & $1024 \times 128 \times 224$ & $50$ & $1.75$ & $1.75$ & $0.00$ &  $   $	&	$   $ &	$   $ & $0.035$ & $0.0375$ \\
\hline
    $0.52$ & $1024 \times 128 \times 192$ & $50$ & $1.75$ & $1.75$ & $0.00$ &  $   $	&	$   $ &	$   $ & $0.03$ & $0.0325$ \\
    $0.52$ & $1024 \times 128 \times 192$ & $50$ & $1.05$ & $1.75$ & $0.69$ &  $5-50 \times 10^2$	& $0.2-2.0$ & $10^2$ & $0.03-0.035$ & $0.0325-0.0375$ \\
\hline
    $0.63$ & $1024 \times 128 \times 160$ & $50$ & $1.75$ & $1.75$ & $0.00$ &  $   $	&	$   $ &	$   $ & $0.025$ & $0.0275$ \\
    $0.63$ & $1024 \times 128 \times 160$ & $50$ & $1.05$ & $1.75$ & $0.69$ &  $10-50 \times 10^2$	& $0.4-2.0$ & $10^2$ & $0.0275-0.04$ & $0.03-0.0425$ \\
\hline
    $0.70$ & $1024 \times 128 \times 160$ & $56$ & $1.75$ & $1.75$ & $0.00$ &  $   $	&	$   $ &	$   $ & $0.0275$ & $0.03$ \\
    $0.70$ & $1024 \times 128 \times 160$ & $56$ & $1.05$ & $1.75$ & $0.69$ &  $5-50 \times 10^2$	& $0.2-2.0$ & $10^2$ & $0.0275-0.0425$ & $0.03-0.045$ \\
\hline
    $0.78$ & $1024 \times 128 \times 128$ & $50$ & $1.75$ & $1.75$ & $0.00$ &  $   $	&	$   $ &	$   $ & $0.0275$ & $0.03$ \\
    $0.78$ & $1024 \times 128 \times 128$ & $50$ & $1.05$ & $1.75$ & $0.69$ &  $2.5-70 \times 10^2$	& $0.1-2.8$ & $10^2$ & $0.0275-0.045$ & $0.03-0.0475$ \\
    $0.78$ & $1024 \times 128 \times 128$ & $50$ & $1.05$ & $1.75$ & $0.69$ &  $2.5-50 \times 10^2$	& $0.1-2.0$ & $10^4$ & $0.02-0.0325$ & $0.0225-0.035$ \\
\hline
    $0.94$ & $1192 \times 128 \times 128$ & $60$ & $1.75$ & $1.75$ & $0.00$ &  $   $	&	$   $ &	$   $ & $0.025$ & $0.0275$ \\
    $0.94$ & $1360 \times 128 \times 128$ & $60$ & $1.05$ & $1.75$ & $0.69$ &  $5-50 \times 10^2$	& $0.2-2.0$ & $10^2$ & $0.025-0.0375$ & $0.0275-0.04$ \\
\hline
   \end{tabular}
   \begin{tabular}{@{\extracolsep{\fill}} |c|c|c|c|c|c|c|c|c|c|c|c|c|}
\hline
    $2R/H$ & $L_{x} \times L_y \times H$ & $R$ &  $\eta_A$ & $\eta_B$  & $\eta_P$ & $\tau_P$ & $De$ & $L^2$ & $U_{\rm w}$ \\
   & cells & lbu & lbu & lbu & lbu & lbu & & & lbu \\
\hline \hline
    $0.78$ & $1024 \times 128 \times 128$ & $50$ & $1.05$ & $1.75$ & $0.69$ &  $~~~~~50 \times 10^2~~~~~$	& $~~~2.0~~~$ &	$10^2, 10^3, 5 \times 10^3, 10^4$ & $~~~~~~~~~~~~0.02~~~~~~~~~~~~$ \\
\hline

\end{tabular}
\caption{\small
Parameters for break-up simulations : $2R/H$ is the confinement ratio, $L_{x}  \times L_y \times H$ is the computational domain, $R$ is the droplet radius, $\eta_A$ is the dynamic viscosity of the Newtonian solvent fluid inside the droplet, $\eta_B$ is the dynamic viscosity of the Newtonian matrix (see also Fig.~\ref{fig:1}), $\eta_P$ is the dynamic viscosity of the polymers, $\tau_P$ is the polymer relaxation time, $L^2$ is the finite extensibility parameter for the polymers (i.e. their maximum squared elongation), $\vpre$ and $\vpost$ are pre-break-up (pre-critical) and post-break-up (post-critical) wall velocity, respectively. \label{table:para}}
\end{table} 

\begin{figure}[pth!]
\subfigure[\,\,t/$\tau_{\mbox{\tiny{em}}}$=25, 2R/H = 0.4, De=0]
    {
        \includegraphics[scale=0.04]{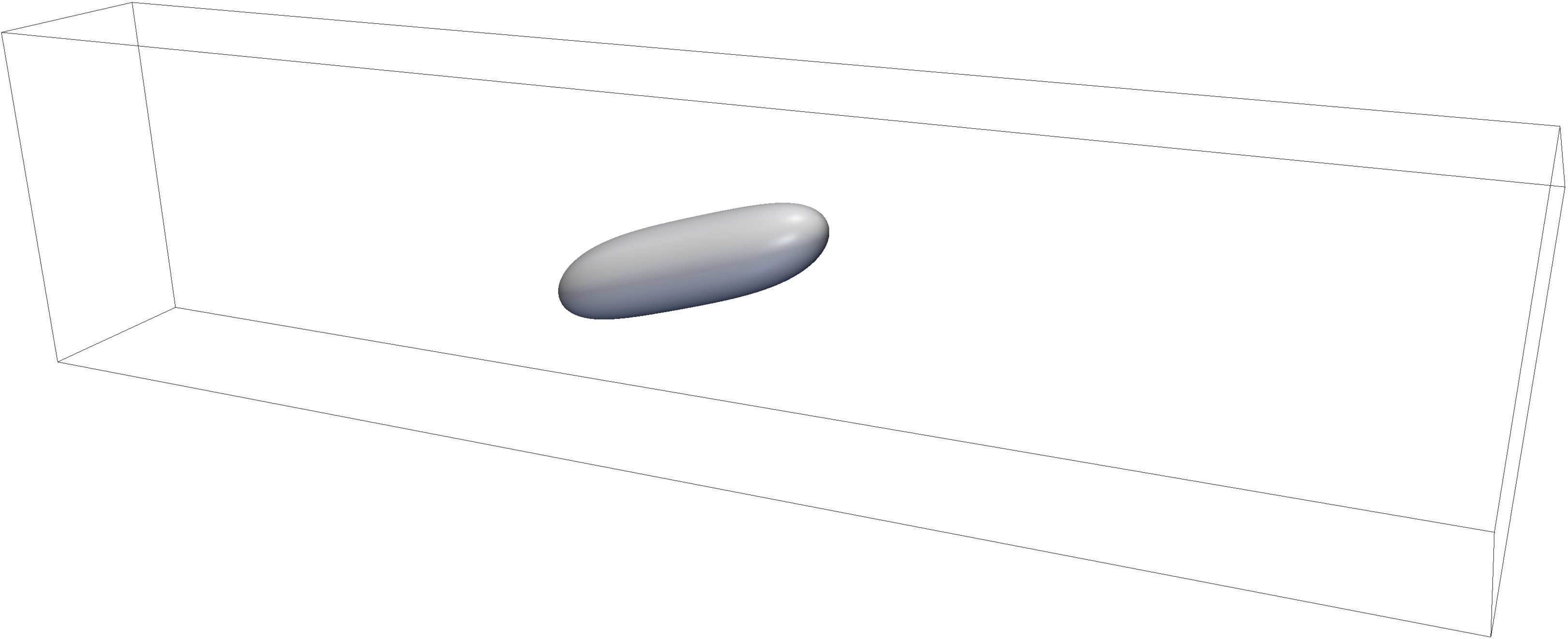}
    }    
\subfigure[\,\,t/$\tau_{\mbox{\tiny{em}}}$=75, 2R/H = 0.4, De=0]
    {
        \includegraphics[scale=0.04]{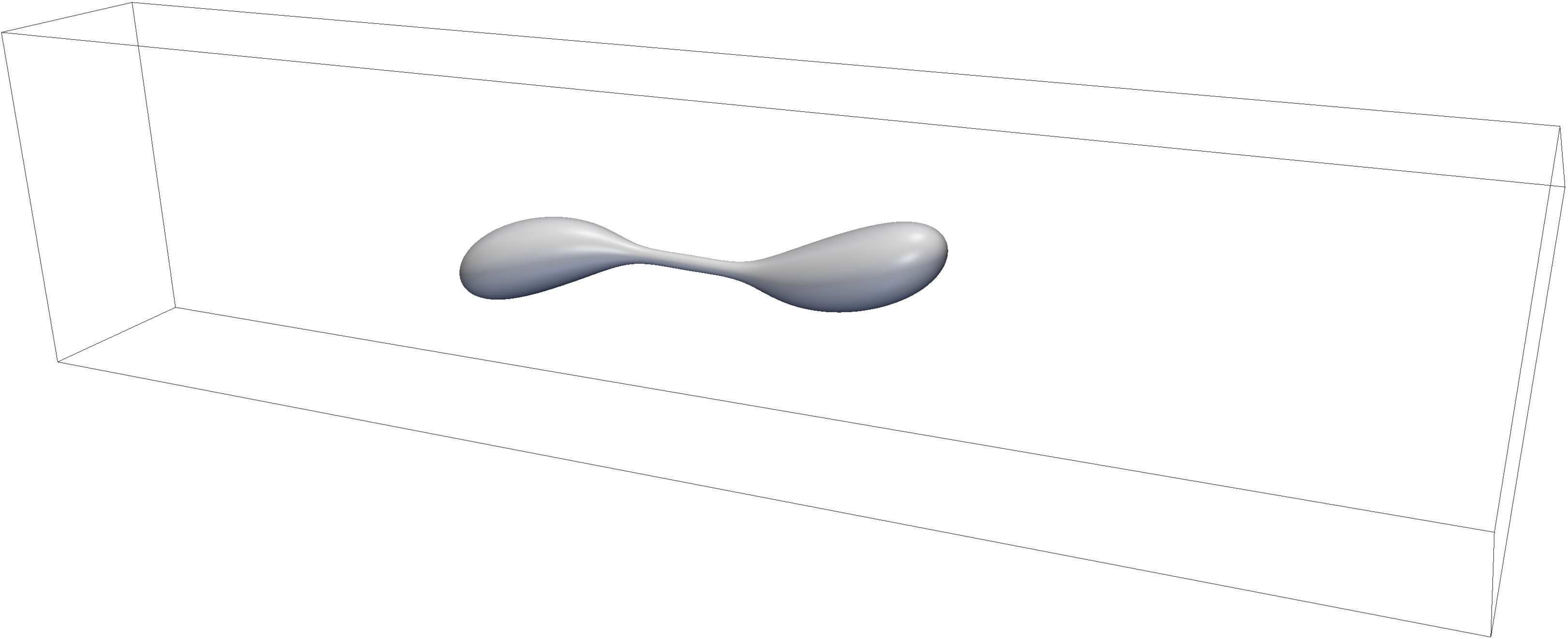}
    }
\subfigure[\,\,t/$\tau_{\mbox{\tiny{em}}}$=100, 2R/H = 0.4, De=0]
    {
        \includegraphics[scale=0.04]{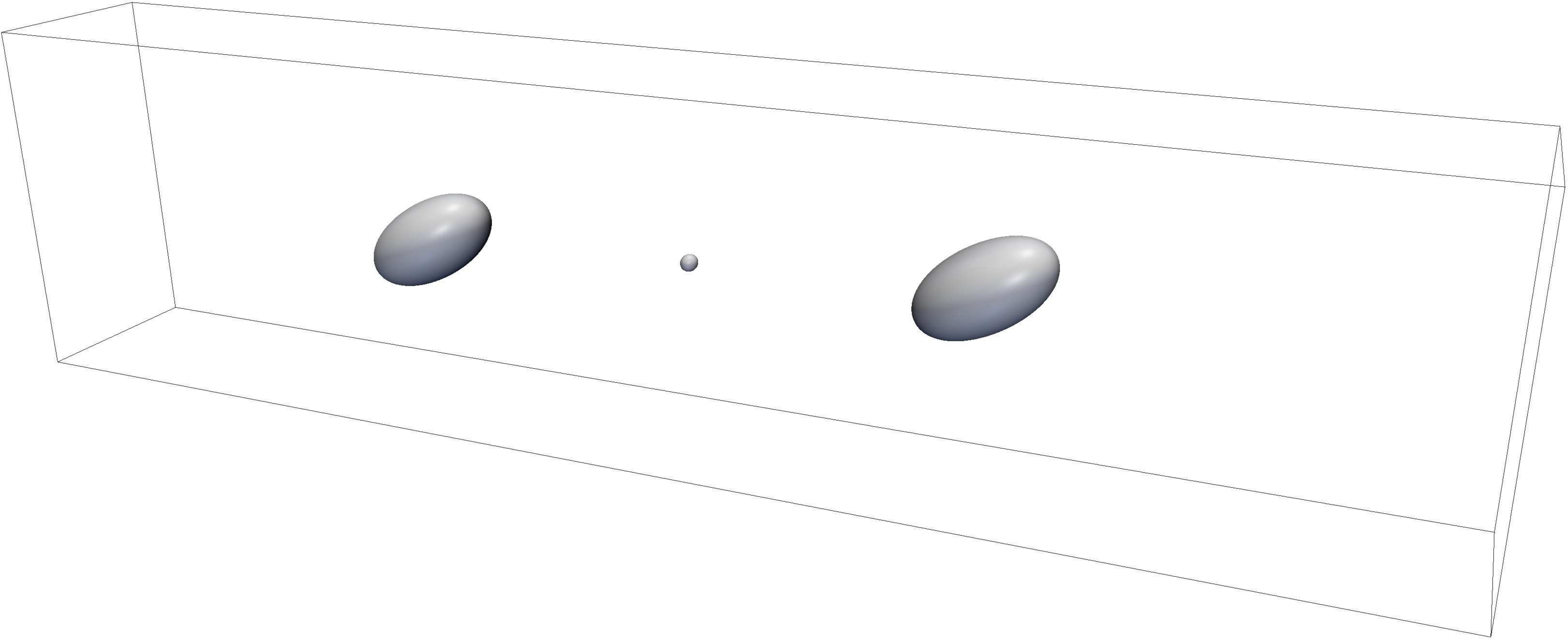}
    }    
\\
\subfigure[\,\,t/$\tau_{\mbox{\tiny{em}}}$=25, 2R/H = 0.4, De=0.20]
    {
        \includegraphics[scale=0.04]{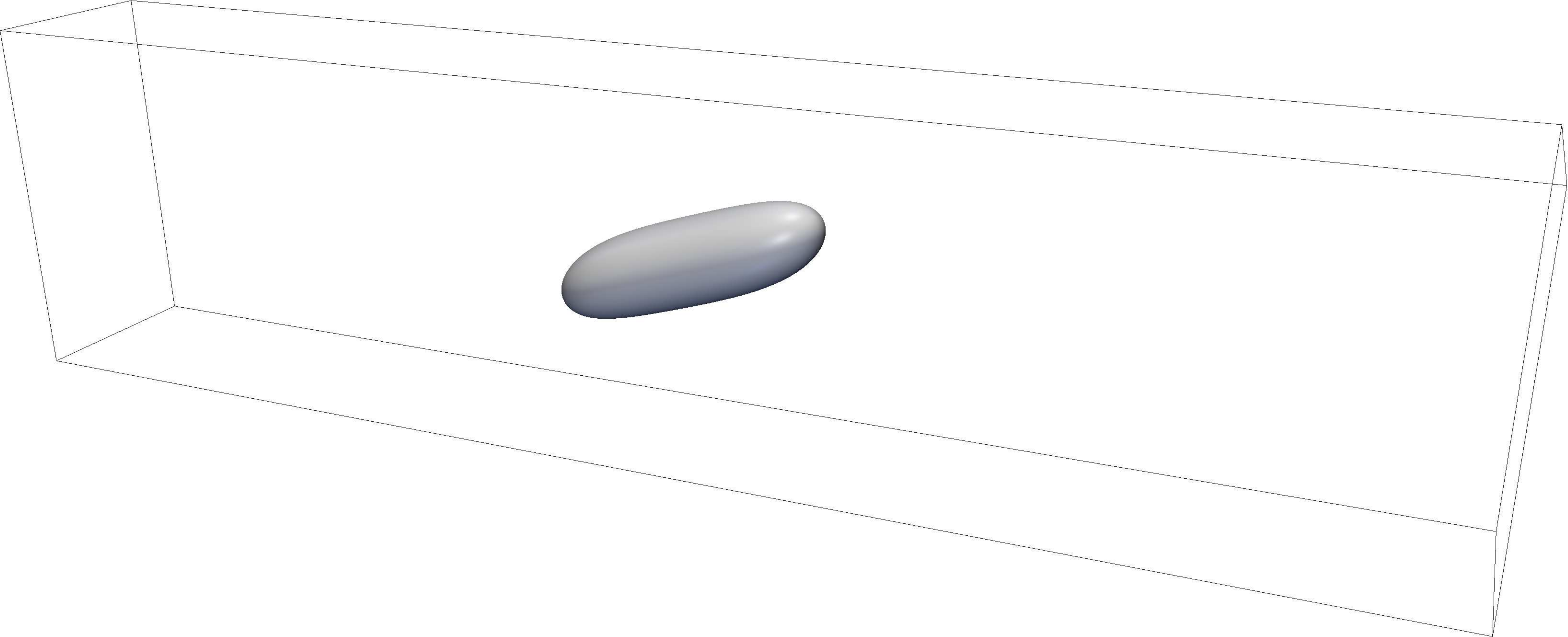}
    }    
\subfigure[\,\,t/$\tau_{\mbox{\tiny{em}}}$=75, 2R/H = 0.4, De=0.20]
    {
        \includegraphics[scale=0.04]{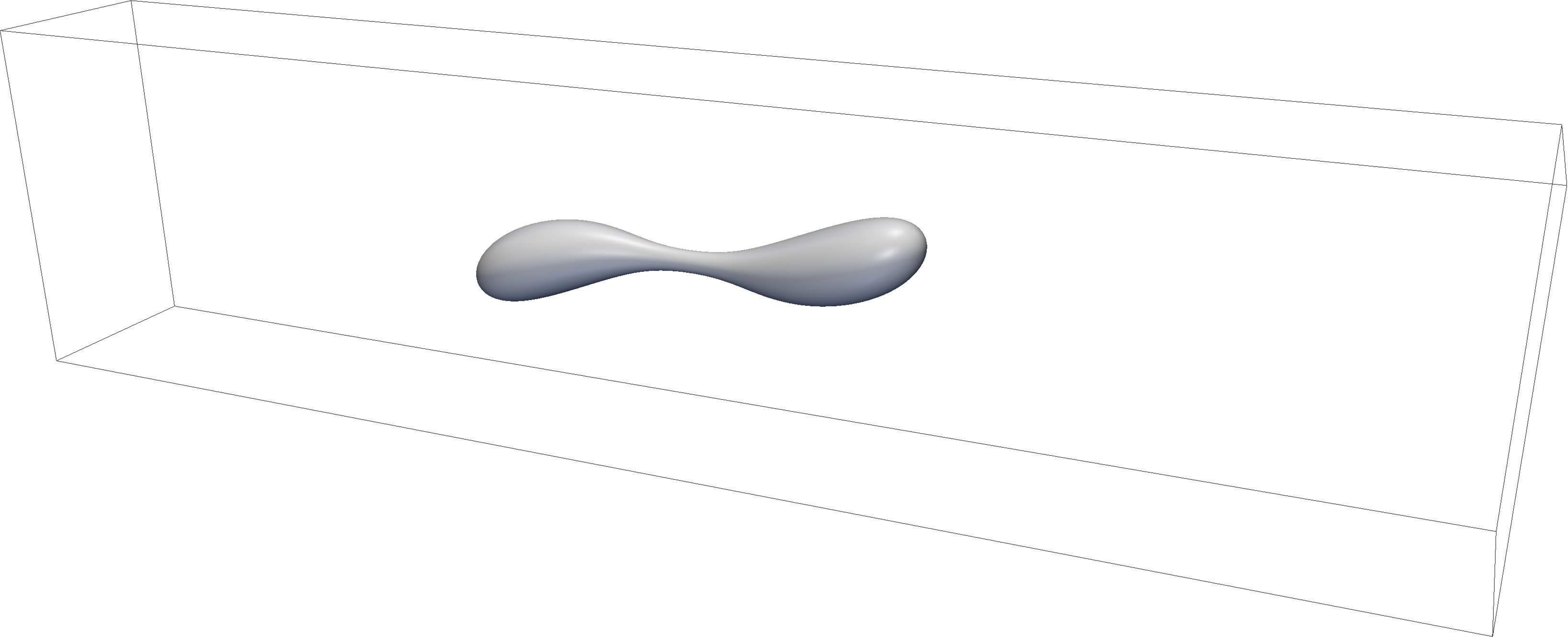}
    }    
\subfigure[\,\,t/$\tau_{\mbox{\tiny{em}}}$=100, 2R/H = 0.4, De=0.20]
    {
        \includegraphics[scale=0.04]{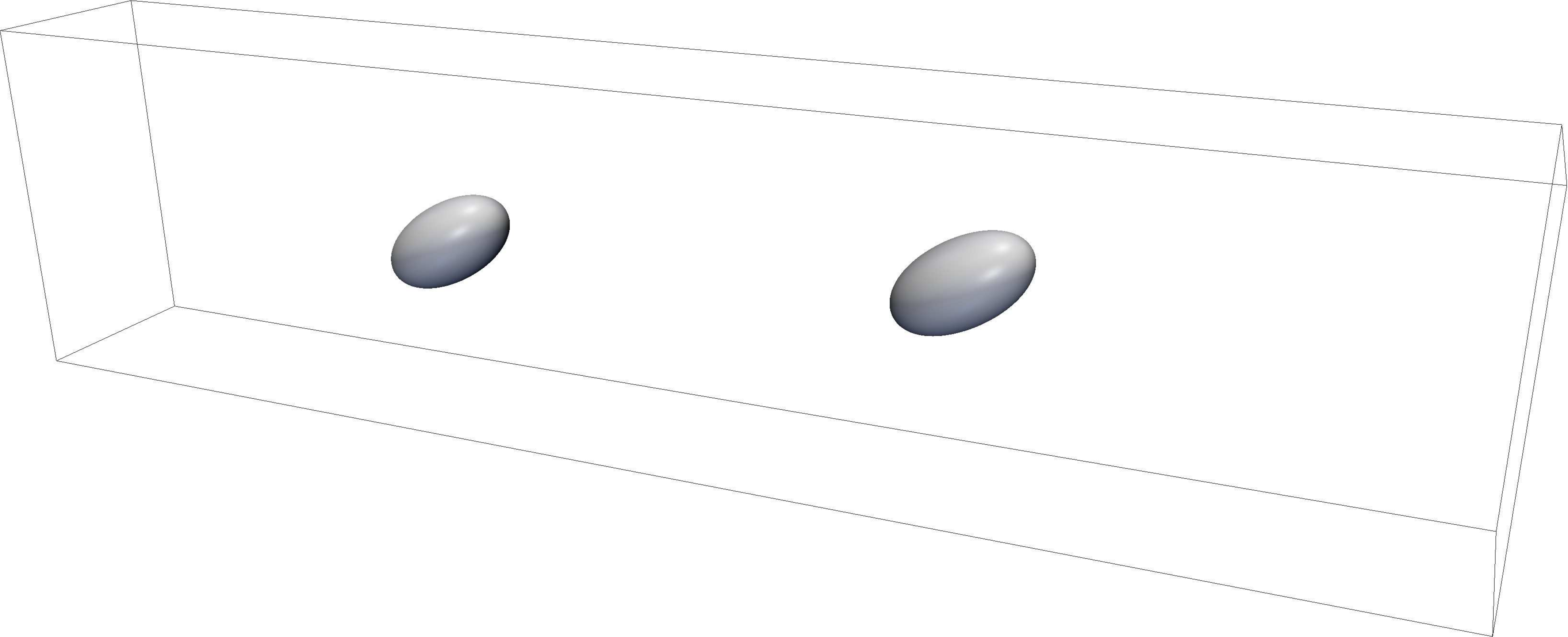}
    }
\\
\subfigure[\,\,t/$\tau_{\mbox{\tiny{em}}}$=25, 2R/H = 0.4, De=2.0]
    {
        \includegraphics[scale=0.04]{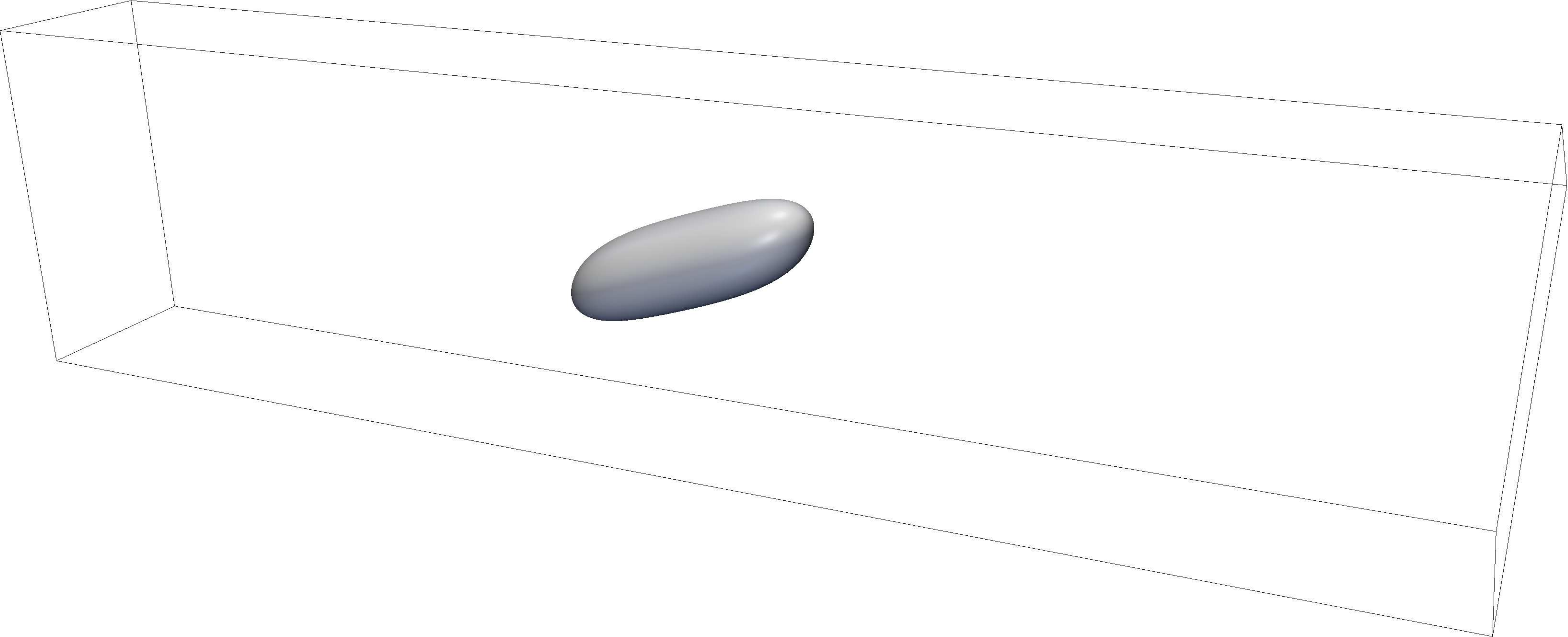}
    }    
\subfigure[\,\,t/$\tau_{\mbox{\tiny{em}}}$=75, 2R/H = 0.4, De=2.0]
    {
        \includegraphics[scale=0.04]{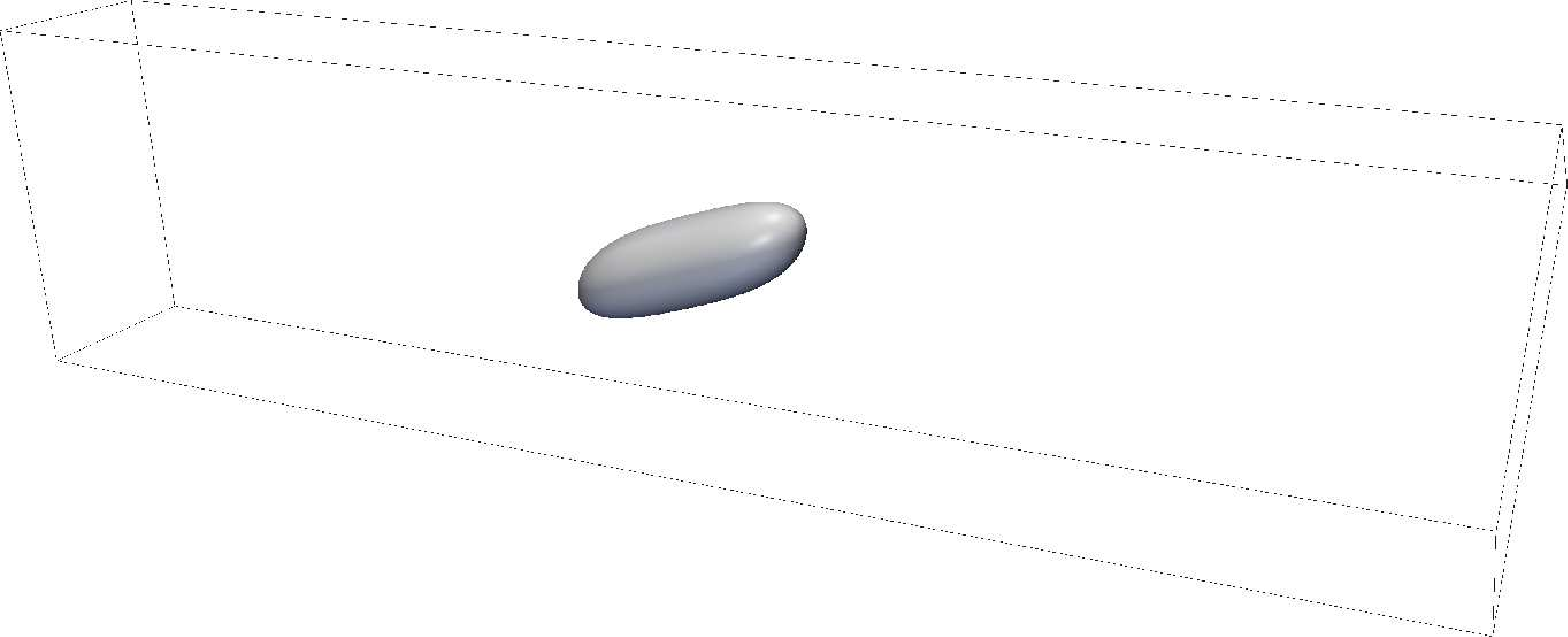}
    }    
\subfigure[\,\,t/$\tau_{\mbox{\tiny{em}}}$=100, 2R/H = 0.4, De=2.0]
    {
        \includegraphics[scale=0.04]{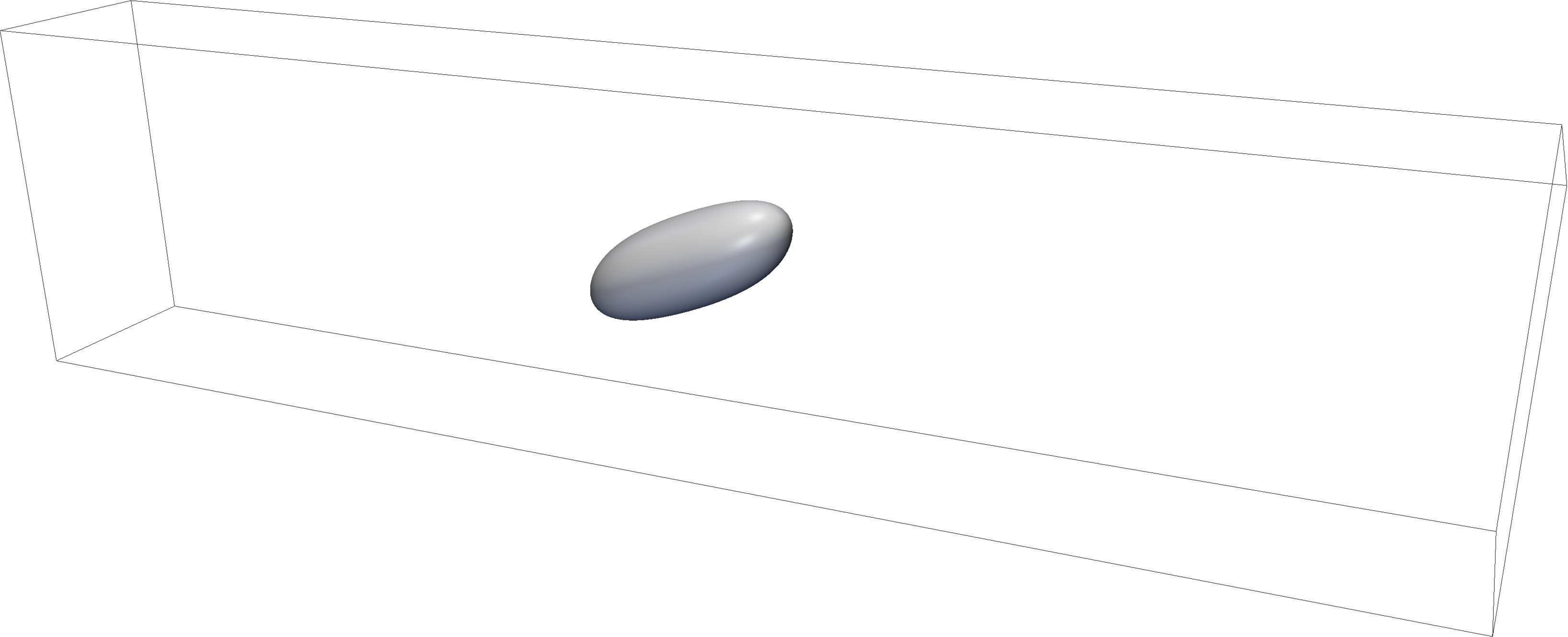}
    }
\caption{Deformation/Break-up after the startup of a shear flow with confinement ratio $2R/H = 0.4$. We report the time history of droplet deformation and break-up including 3 time frames which represent, in the Newtonian case ($\mbox{De}=0$), initial deformation (left column, $t= 25 \tau_{\mbox{\tiny{em}}}$); deformation prior to break-up (middle column, $t = 75 \tau_{\mbox{\tiny{em}}}$); post-break-up frame (right column, $t = 100\tau_{\mbox{\tiny{em}}}$). We use the emulsion time $\tau_{\mbox{\tiny{em}}}$ (see Eq.~\eqref{emulsiontime}) as a unit of time. The second row of images is related to a weakly viscoelastic droplet ($\mbox{De}=0.2$), indicating that non-Newtonian properties do not affect the droplet deformation and break-up much. The third row of images is related to a viscoelastic droplet with Deborah number above unity ($\mbox{De}=2.0$), and indicates that non-Newtonian properties stabilize the droplet deformation and inhibit droplet break-up. Note that the Capillary number is kept fixed to the post-critical Newtonian value $\mbox{Ca}=0.34$,  which is the smallest Capillary number available for us at which we observe break-up in the Newtonian case. In all cases, the viscous ratio between the droplet phase and the matrix phase is kept fixed to $\lambda=\eta_D/\eta_M=1$, independently of the degree of viscoelasticity. The finite extensibility parameter is fixed to $L^2=10^2$. } \label{fig:3}
\end{figure}

\begin{figure}[pth!]
\subfigure[\,\,t/$\tau_{\mbox{\tiny{em}}}$=25, 2R/H = 0.4, Ca=0.34, De=2.0]
    {
        \includegraphics[scale=0.04]{unbounded_stronglyvisco_t_25000}
    }    
\subfigure[\,\,t/$\tau_{\mbox{\tiny{em}}}$=75, 2R/H = 0.4, Ca=0.34, De=2.0]
    {
        \includegraphics[scale=0.04]{unbounded_stronglyvisco_t_75000}
    }    
\subfigure[\,\,t/$\tau_{\mbox{\tiny{em}}}$=100, 2R/H = 0.4, Ca=0.34, De=2.0]
    {
        \includegraphics[scale=0.04]{unbounded_stronglyvisco_t_100000}
    }
\\
\subfigure[\,\,t/$\tau_{\mbox{\tiny{em}}}$=25, 2R/H = 0.4, Ca=0.35, De=2.0]
    {
        \includegraphics[scale=0.04]{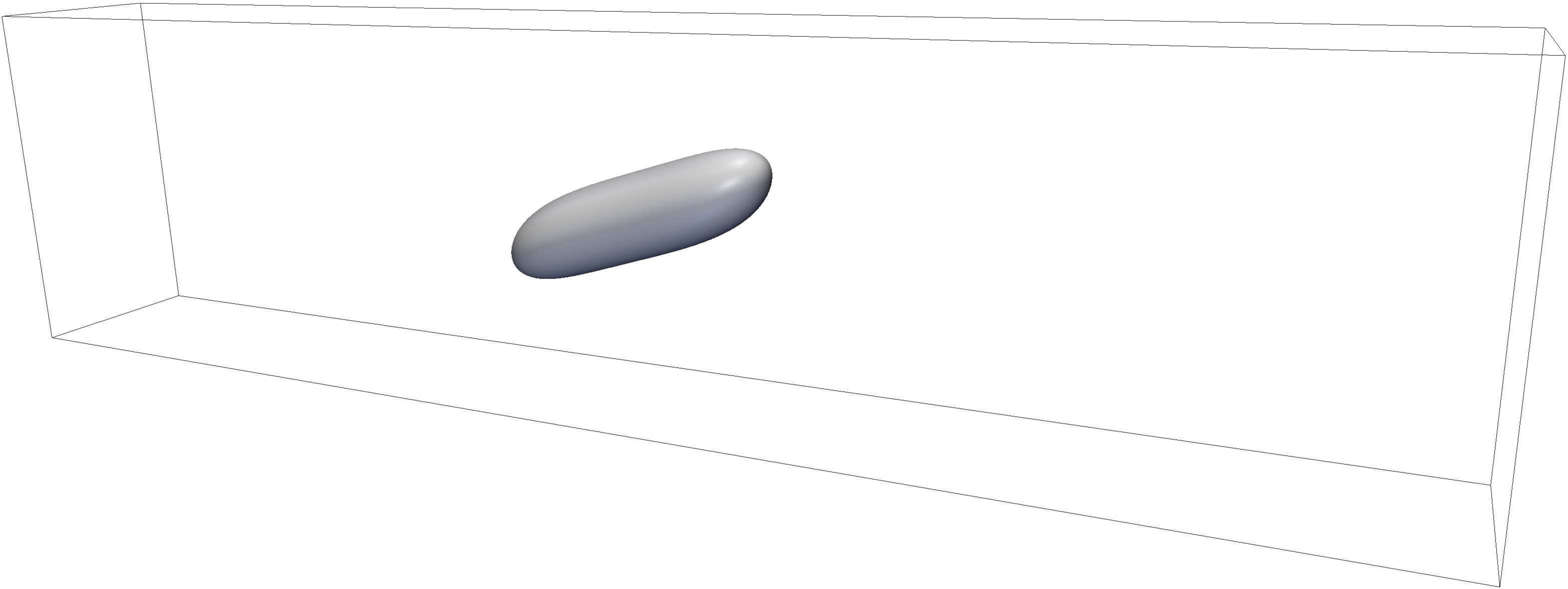}
    }    
\subfigure[\,\,t/$\tau_{\mbox{\tiny{em}}}$=75, 2R/H = 0.4, Ca=0.35, De=2.0]
    {
        \includegraphics[scale=0.04]{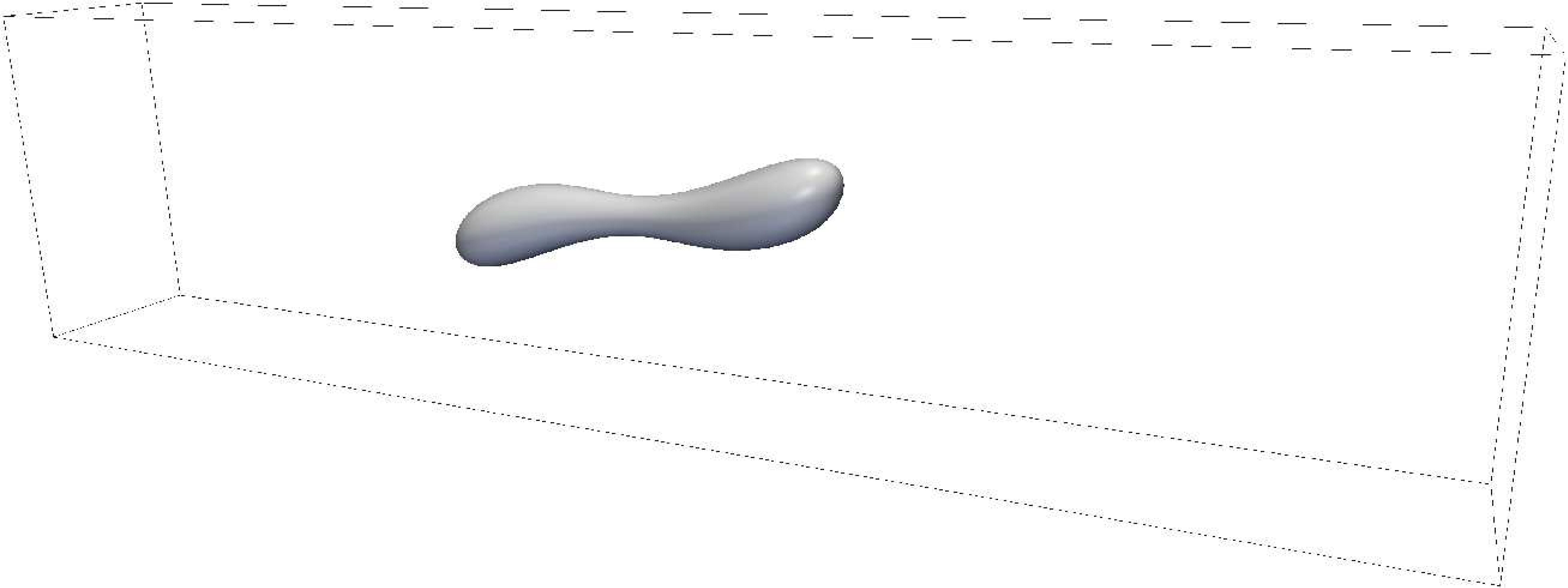}
    }    
\subfigure[\,\,t/$\tau_{\mbox{\tiny{em}}}$=100, 2R/H = 0.4, Ca=0.35, De=2.0]
    {
        \includegraphics[scale=0.04]{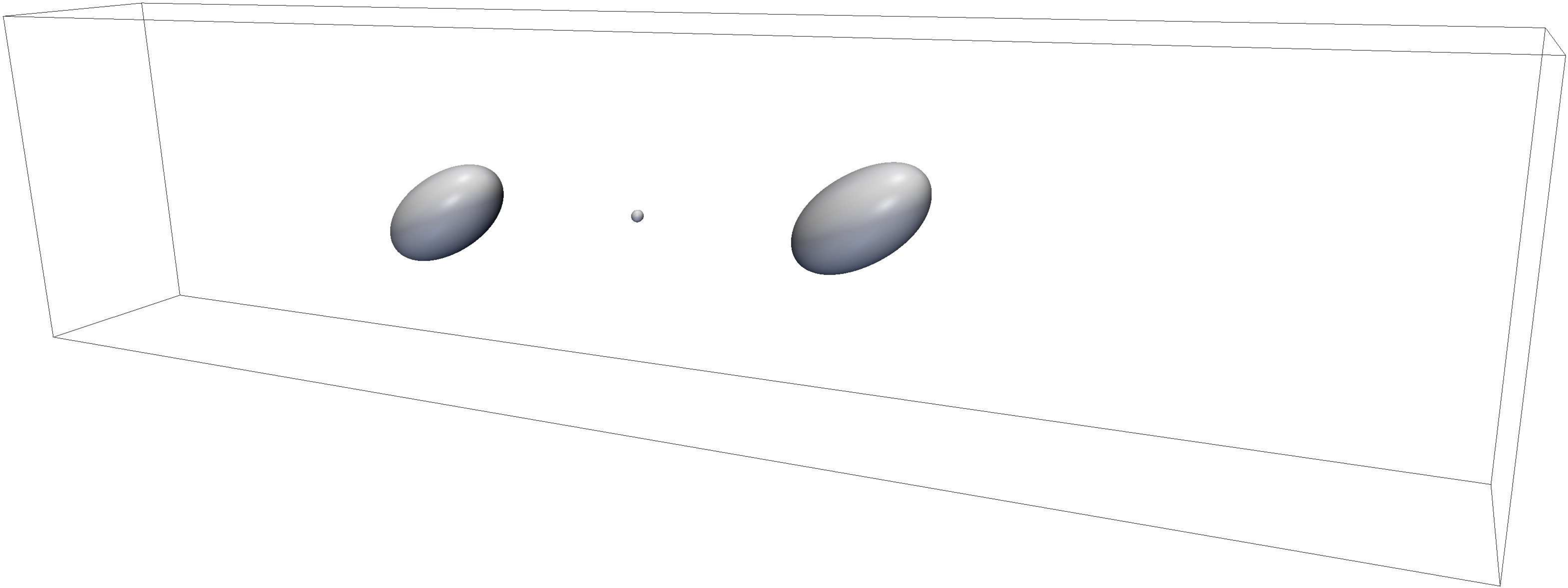}
    }
\caption{Deformation/Break-up of a viscoelastic droplet in a Newtonian matrix after the startup of a shear flow with confinement ratio $2R/H = 0.4$ and Deborah number $\mbox{De}=2.0$, at changing the Capillary number.  The finite extensibility parameter is fixed to $L^2=10^2$. The first row of images is just the last row of images in Fig.~\ref{fig:3}, corresponding to $\mbox{Ca}= 0.34$. The droplet deformation increases with increasing $\mbox{Ca}$, and when $\mbox{Ca}$ exceeds a critical value $\mbox{Ca}_{\mbox{\tiny{cr}}}$ between $0.34$ and $0.35$ the droplet breaks into two equally sized droplets (second row of images). The critical Capillary number at $\mbox{De}=2.0$ is close to the Newtonian counterpart ($\mbox{De}=0$, see Fig.~\ref{fig:3}). In all cases, the viscous ratio between the droplet phase and the matrix phase is kept fixed to $\lambda=\eta_D/\eta_M=1$, independently of the degree of viscoelasticity. \label{fig:4}}
\end{figure}

In Fig.~\ref{fig:3} we report 3D snapshots showing deformation and subsequent break-up of the droplet after the startup of a shear flow with the smallest confinement ratio analyzed in our numerical simulations, $2R/H = 0.4$, at fixed Capillary number. The Capillary number is chosen to be the critical Capillary number for the Newtonian droplet ($\mbox{Ca}=0.34$). Panels (a)-(c) refer to the Newtonian case and they show the initial droplet deformation at time $t = 25 \tau_{\mbox{\tiny{em}}}$, the droplet deformation prior to break-up at time $t = 75 \tau_{\mbox{\tiny{em}}}$, and the droplet in post-break-up conditions at $t = 100 \tau_{\mbox{\tiny{em}}}$, respectively. Panels (d)-(f)  and panels (g)-(i) show the behavior at changing the Deborah number, obtained by changing the relaxation time $\tau_P$ in Eqs.\eqref{NS}-\eqref{FENE}. Panels (d)-(f) show the results for a slightly viscoelastic case ($\mbox{De} = 0.2$). Clearly, in presence of weak viscoelastic effects, the droplet dynamics is very close to the Newtonian case, with little resistance against deformation. Panels (g)-(i) show the results for a viscoelastic case with Deborah number above unity ($\mbox{De} = 2.0$). In this case, the viscoelastic droplet does not break, indicating that viscoelasticity has a stabilizing effect on the droplet and prevents the droplet break-up. However, this stabilization is not remarkable, since a slight increase in $\mbox{Ca}$ leads to droplet break-up. This is shown in Fig.~\ref{fig:4}, where the last row of images of Fig.~\ref{fig:3} is compared with the corresponding images at a slightly larger Capillary number, $\mbox{Ca} \approx 0.35$. We remark that we set our parameters in such a way that the viscous ratio between the droplet phase and the matrix phase is kept fixed to $\lambda=1$. This is done to appreciate in full the role of non-Newtonian effects, which seem to be rather small at this stage of the analysis. Some words of caution for the values of the critical Capillary numbers studied are also in order. We notice that the critical Capillary number for the Newtonian case ($\mbox{De} = 0$) is found to be $\mbox{Ca}_{\mbox{\tiny{cr}}} = 0.34$, which is different from the usual unconfined result $\mbox{Ca}_{\mbox{\tiny{cr}}}=0.43$~\cite{Grace,Janssen10}. We attribute this difference to the finite Reynolds number of our simulations, which is close to $\mbox{Re}=0.1$. Indeed, Renardy \& Cristini~\cite{RenardyCristini01} in their numerical study using a volume-of-fluid (VOF) method, determined the critical Capillary number at $\mbox{Re} = 0.1$ and $\lambda=1$ to be  $\mbox{Ca}_{\mbox{\tiny{cr}}} \approx 0.38$, which is well in agreement with our finding.  


\begin{figure}[pth!]
\subfigure[\,\,t/$\tau_{\mbox{\tiny{em}}}$=25, 2R/H = 0.78, De=0]
    {
        \includegraphics[scale=0.036]{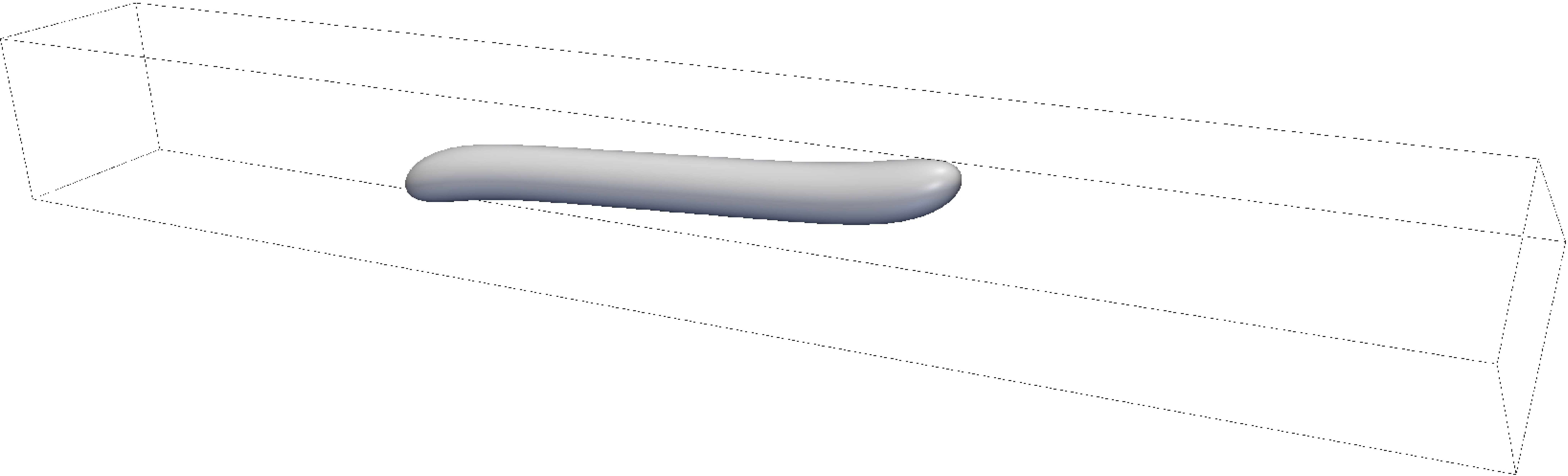}
    }    
\subfigure[\,\,t/$\tau_{\mbox{\tiny{em}}}$=75, 2R/H = 0.78, De=0]
    {
        \includegraphics[scale=0.036]{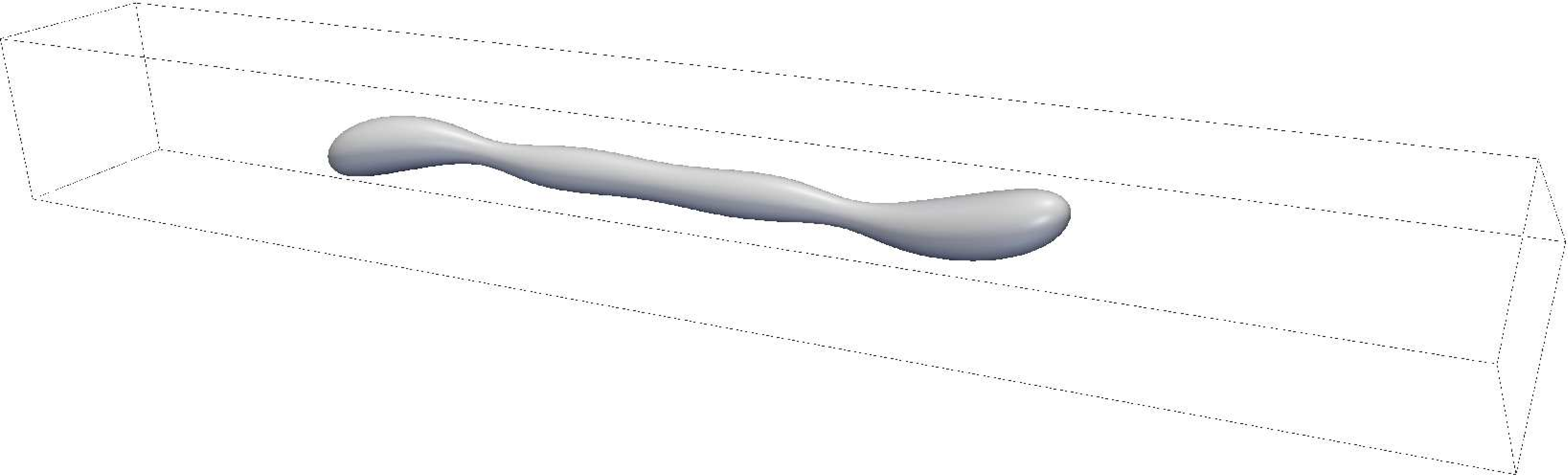}
    }
\subfigure[\,\,t/$\tau_{\mbox{\tiny{em}}}$=100, 2R/H = 0.78, De=0]
    {
        \includegraphics[scale=0.036]{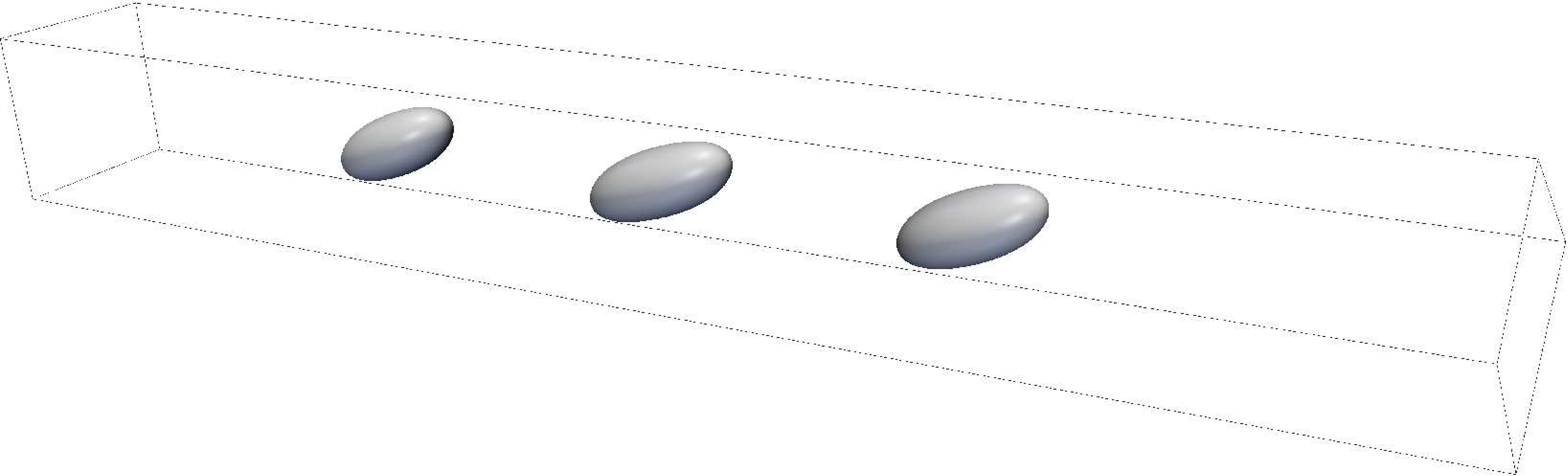}

    }    
\\
\subfigure[\,\,t/$\tau_{\mbox{\tiny{em}}}$=25, 2R/H = 0.78, De=0.2]
    {
        \includegraphics[scale=0.036]{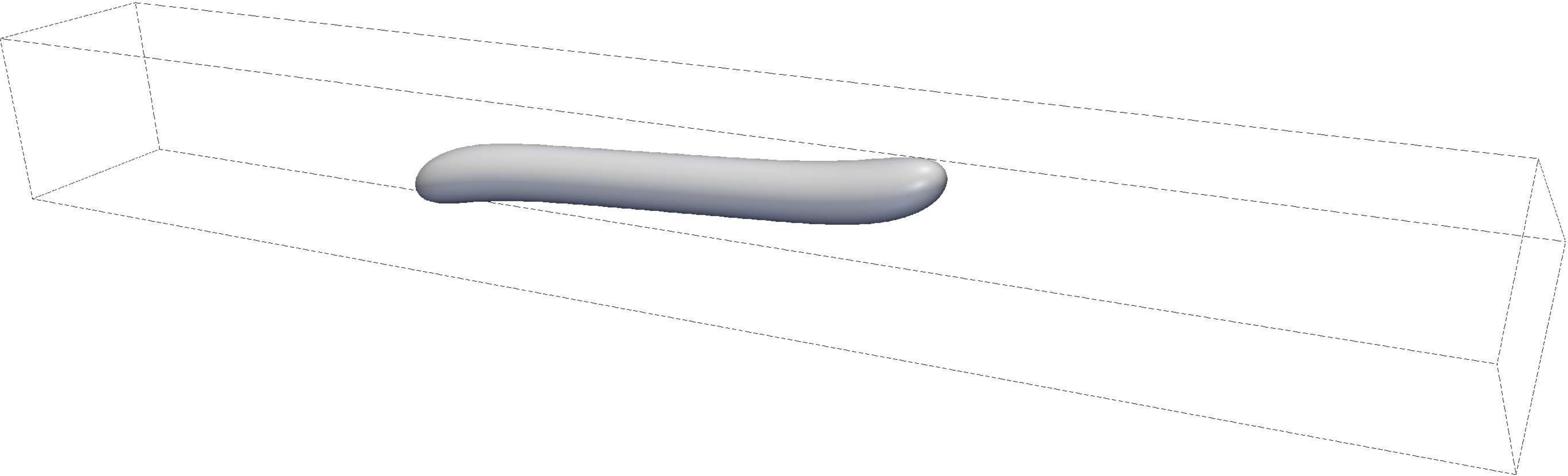}
    }    
\subfigure[\,\,t/$\tau_{\mbox{\tiny{em}}}$=75, 2R/H = 0.78, De=0.2]
    {
        \includegraphics[scale=0.036]{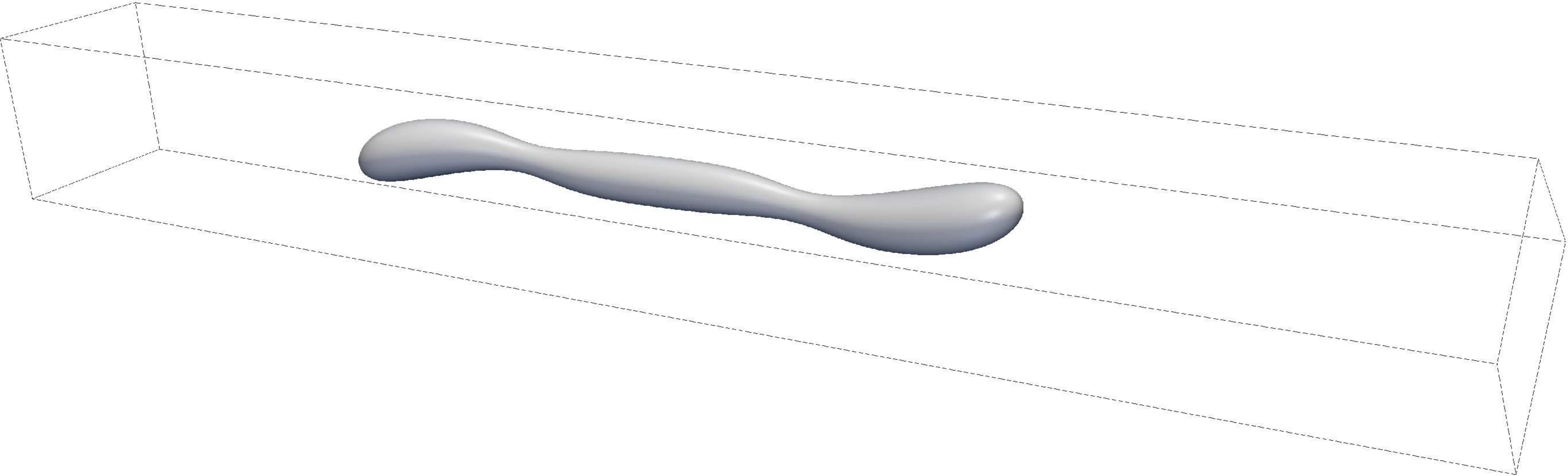}
    }    
\subfigure[\,\,t/$\tau_{\mbox{\tiny{em}}}$=100, 2R/H = 0.78, De=0.2]
    {
        \includegraphics[scale=0.036]{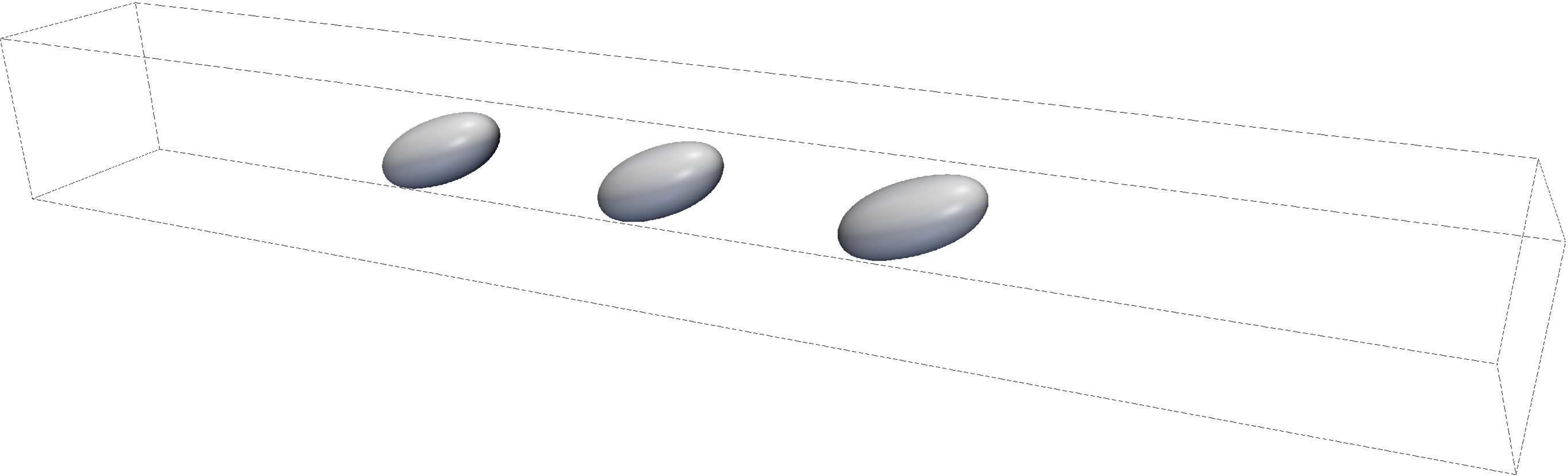}
    }
\\
\subfigure[\,\,t/$\tau_{\mbox{\tiny{em}}}$=25, 2R/H = 0.78, De=2.0]
    {
        \includegraphics[scale=0.036]{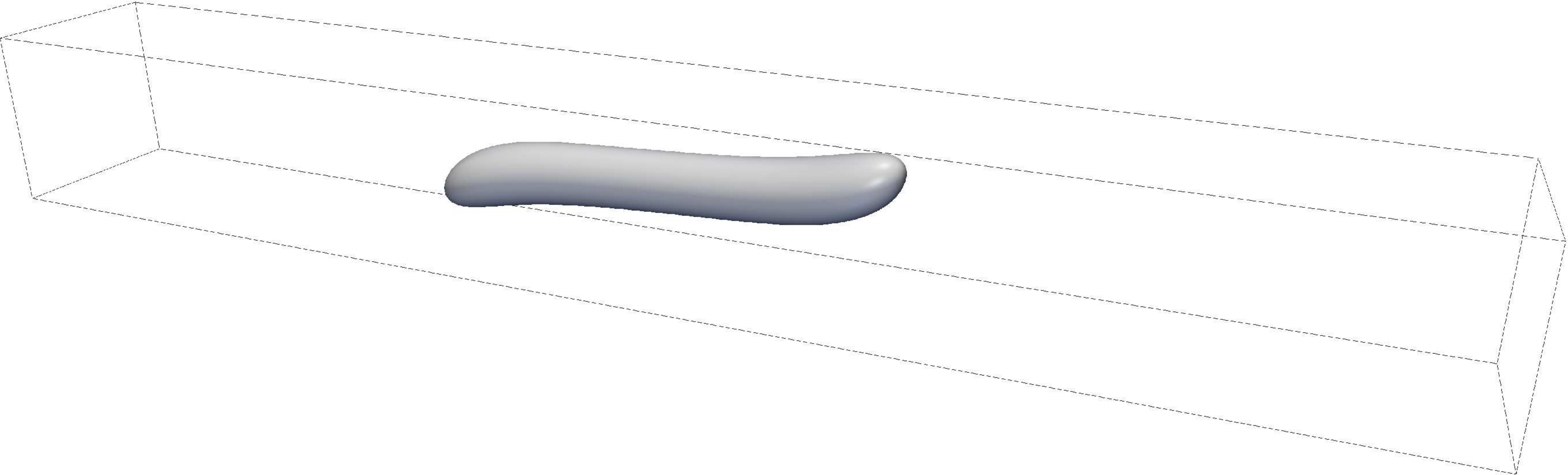}
    }    
\subfigure[\,\,t/$\tau_{\mbox{\tiny{em}}}$=75, 2R/H = 0.78, De=2.0]
    {
        \includegraphics[scale=0.036]{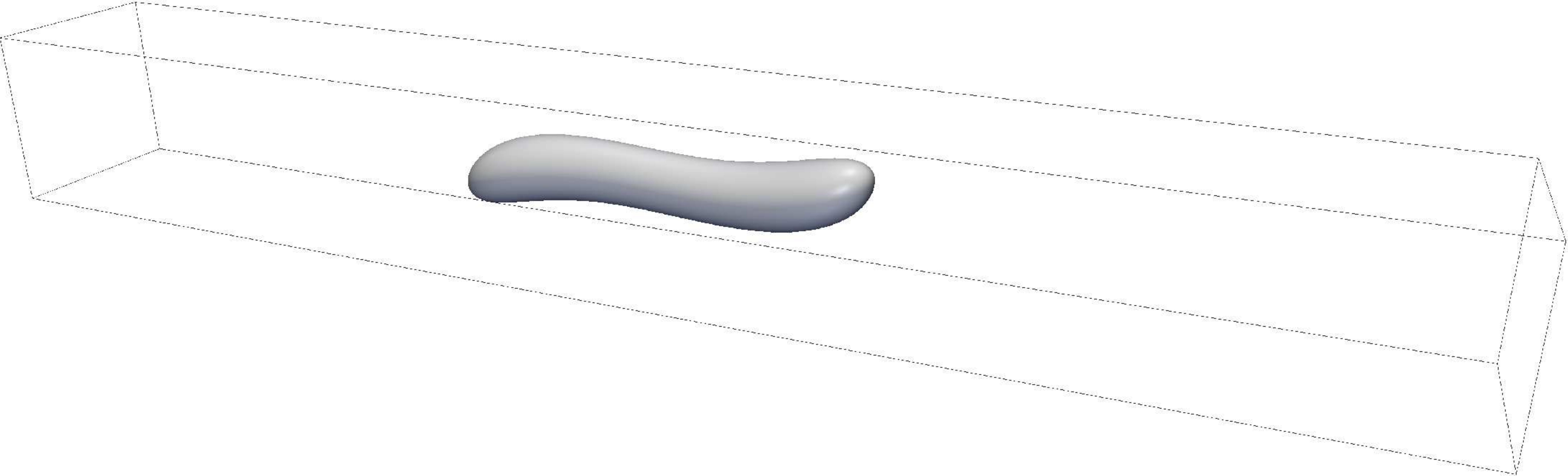}
    }
\subfigure[\,\,t/$\tau_{\mbox{\tiny{em}}}$=100, 2R/H = 0.78, De=2.0]
    {
        \includegraphics[scale=0.036]{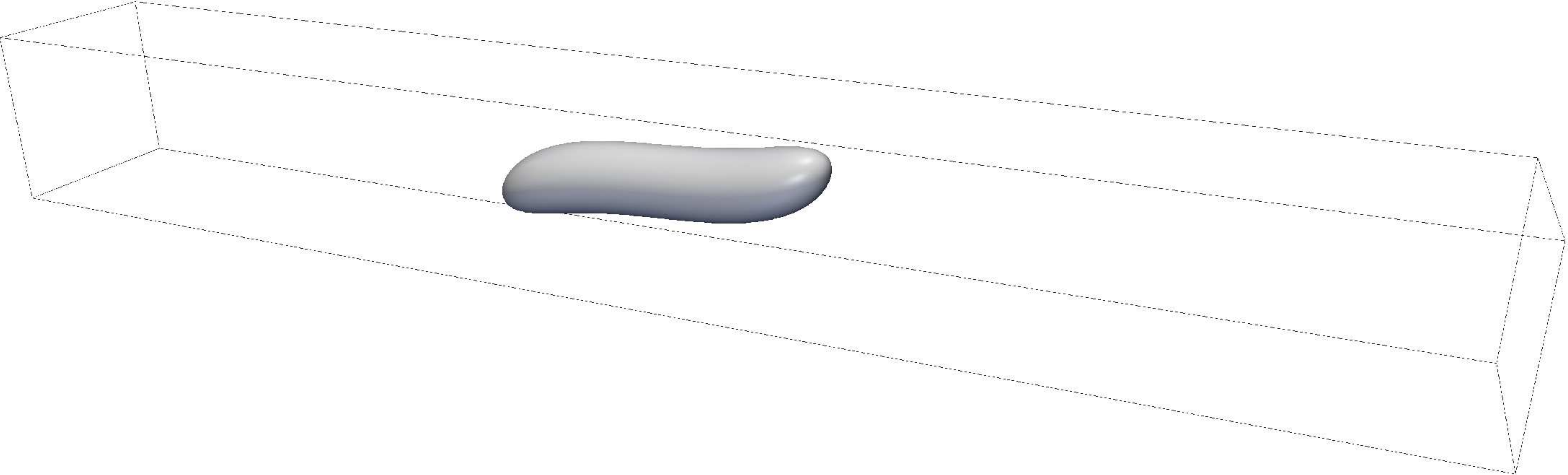}
    }
\caption{Deformation/Break-up after the startup of a shear flow with confinement ratio $2R/H = 0.78$. We report the time history of droplet deformation and break-up including 3 representative time frames, similarly to what is reported in Fig.~\ref{fig:3}. A distinctive feature of this confined case is the emergence of {\it triple break-up}~\cite{Janssen10}. The second row of images is related to a weekly viscoelastic droplet ($\mbox{De}=0.2$), indicating that non-Newtonian properties do not affect the droplet deformation and break-up much. The third row of images is related to a viscoelastic droplet with Deborah number above unity ($\mbox{De}=2.0$), and indicates that such non-Newtonian properties stabilize the droplet deformation and inhibit droplet break-up. Note that the Capillary number is kept fixed to the post-critical Newtonian value $\mbox{Ca}=0.47$, which is the smallest Capillary number at which we observe break-up in the Newtonian case.  In all cases, the viscous ratio between the droplet phase and the matrix phase is kept fixed to $\lambda=\eta_D/\eta_M=1$, independently of the degree of viscoelasticity. The finite extensibility parameter is fixed to $L^2=10^2$. \label{fig:5}}
\end{figure}

\begin{figure}[pth!]
\subfigure[\,\,t/$\tau_{\mbox{\tiny{em}}}$=25, 2R/H = 0.78, Ca=0.47, De=2.0]
    {
        \includegraphics[scale=0.036]{confined_stronglyvisco_t_25000}
    }    
\subfigure[\,\,t/$\tau_{\mbox{\tiny{em}}}$=75, 2R/H = 0.78, Ca=0.47, De=2.0]
    {
        \includegraphics[scale=0.036]{confined_stronglyvisco_t_75000}
    }    
\subfigure[\,\,t/$\tau_{\mbox{\tiny{em}}}$=100, 2R/H = 0.78, Ca=0.47, De=2.0]
    {
        \includegraphics[scale=0.036]{confined_stronglyvisco_t_100000}
    }
\\
\subfigure[\,\,t/$\tau_{\mbox{\tiny{em}}}$=25, 2R/H = 0.78, Ca=0.75, De=2.0]
    {
        \includegraphics[scale=0.036]{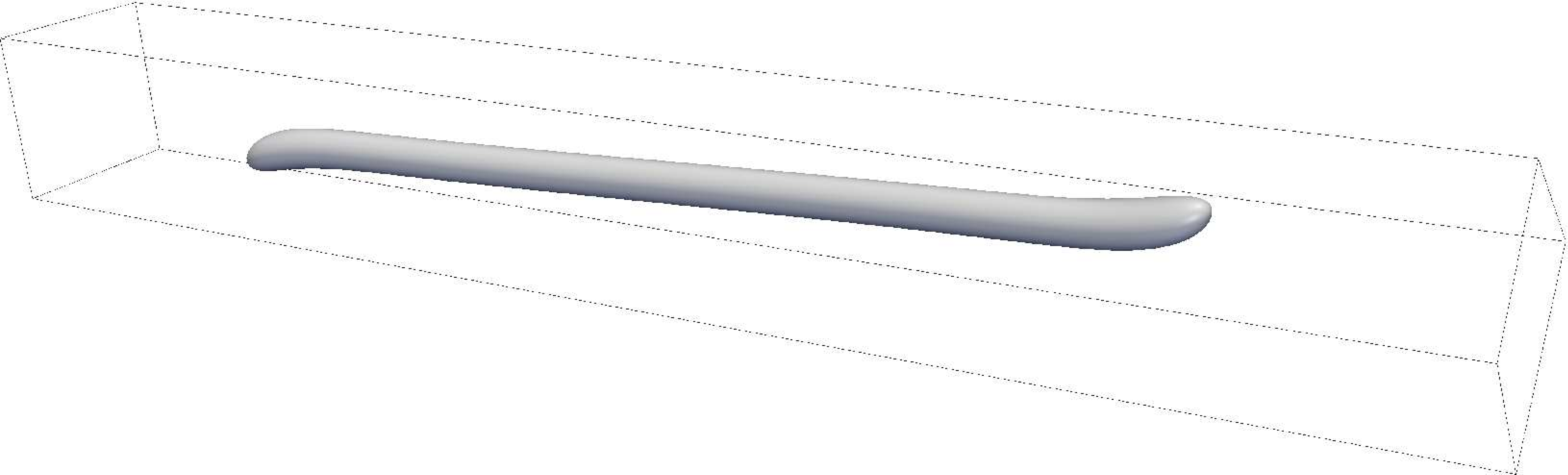}
    }   
\subfigure[\,\,t/$\tau_{\mbox{\tiny{em}}}$=100, 2R/H = 0.78, Ca=0.75, De=2.0]
    {
        \includegraphics[scale=0.036]{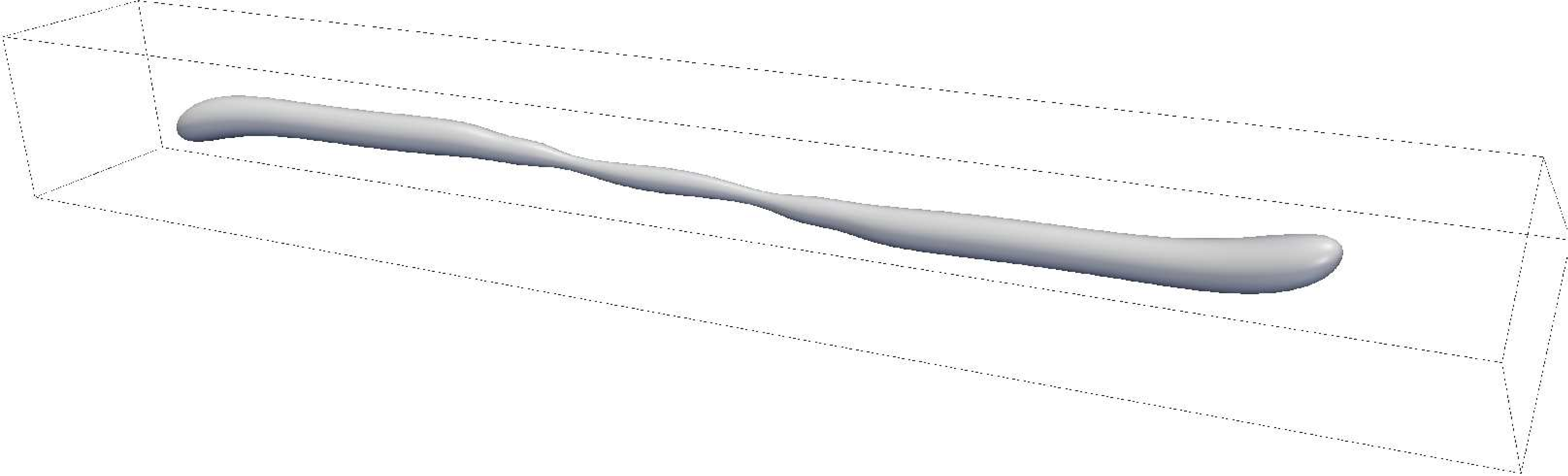}
    }    
\subfigure[\,\,t/$\tau_{\mbox{\tiny{em}}}$=125, 2R/H = 0.78, Ca=0.75, De=2.0]
    {
        \includegraphics[scale=0.036]{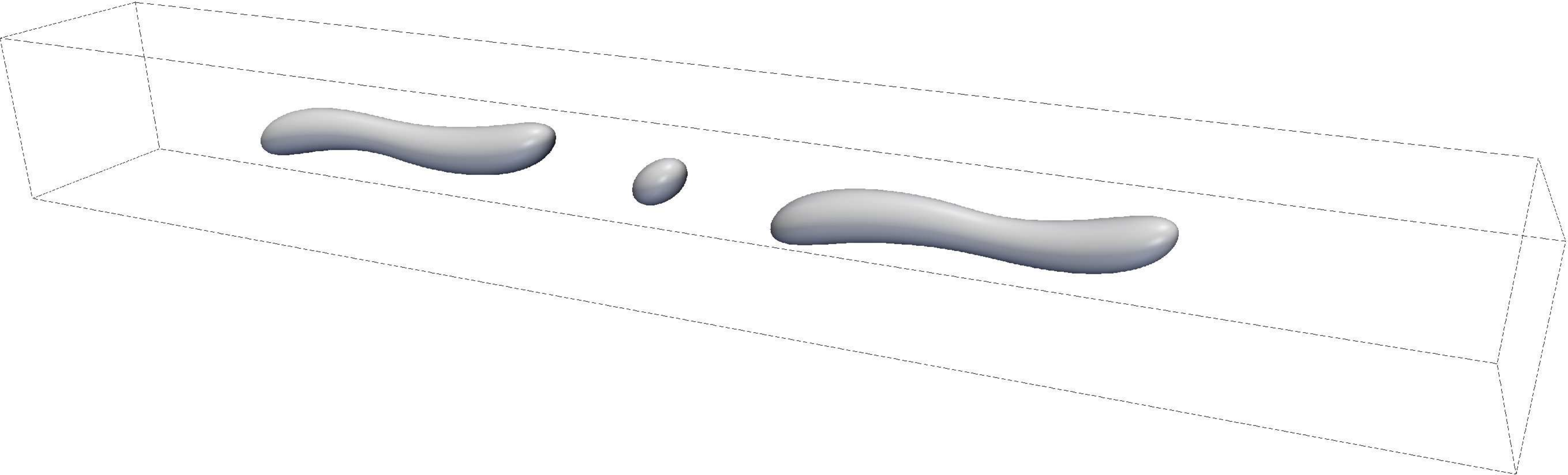}
    }
\caption{Deformation/Break-up of a viscoelastic droplet in a Newtonian matrix after the startup of a shear flow with confinement ratio $2R/H = 0.78$ and Deborah number above unity ($\mbox{De} = 2.0$). The finite extensibility parameter is fixed to $L^2=10^2$. The first row of images is just the last row of images in Fig.~\ref{fig:5}, corresponding to $\mbox{Ca}= 0.47$. The droplet deformation increases with increasing $\mbox{Ca}$ and when $\mbox{Ca}$ exceeds a critical value $\mbox{Ca}_{\mbox{\tiny{cr}}}$ the droplet breaks (second row of images). The critical Capillary number is found to be $\mbox{Ca}_{\mbox{\tiny{cr}}}=0.75$ and increases substantially compared to its Newtonian counterpart ($\mbox{De}=0$, see Fig.~\ref{fig:5}). In all cases, the viscous ratio between the droplet phase and the matrix phase is kept fixed to $\lambda=\eta_D/\eta_M=1$, independently of the degree of viscoelasticity. \label{fig:6}}
\end{figure}

We next perform a similar analysis for a case where the droplet is in a highly confined situation. In Fig.~\ref{fig:5} we show 3D snapshots including deformation and subsequent break-up of the droplet after the startup of a shear flow, with a confinement ratio $2R/H = 0.78$, at fixed Capillary number. Similarly to Fig.~\ref{fig:3}, the first row of images is related to the Newtonian case: panels (a)-(c) show the initial droplet deformation at time $t = 25 \tau_{\mbox{\tiny{em}}}$, the droplet deformation prior to break-up at time $t = 75 \tau_{\mbox{\tiny{em}}}$, and in a post-break-up condition at $t = 100 \tau_{\mbox{\tiny{em}}}$, respectively. It must be noted that confinement acts in stabilizing the droplet with elongated shapes that would be unstable in an unconfined case~\cite{Sibillo06}. Upon elongation, the droplet now breaks into three (more than two) equally sized droplets, due to the Rayleigh-Plateau instability that develops at the interface~\cite{Janssen10,Cardinaels11}.  Compared to the lower confinement ratio analyzed in the first row of Fig.~\ref{fig:3}, the critical Capillary number increases because of the stabilizing effect of the wall and the associated different break-up mechanism. We estimate $\mbox{Ca}_{\mbox{\tiny{cr}}}=0.47$ compared to $\mbox{Ca}_{\mbox{\tiny{cr}}}=0.34$ estimated in the lower confinement ratio.  Panels (d)-(f) and panels (g)-(i) show the behavior at changing the Deborah number, obtained by changing the relaxation time $\tau_P$ in Eqs.~\eqref{NS}-\eqref{FENE}. Similarly to the unconfined case analyzed in Fig.~\ref{fig:3}, viscoelasticity stabilizes the droplet and prevents the droplet break-up. However, a net distinction between the unconfined case ($2R/H = 0.4$) and the confined case ($2R/H = 0.78$) emerges. At fixed Deborah number, break-up in the confined case is observed at a much higher $\mbox{Ca}_{\mbox{\tiny{cr}}}$ than the Newtonian case. This is quantitatively visualized in Fig.~\ref{fig:6}, where we show the history of the deformation and break-up of the viscoelastic droplet for $\mbox{De} = 2.0$ and confinement ratio $2R/H = 0.78$. The critical Capillary number is measured to be $\mbox{Ca}_{\mbox{\tiny{cr}}}=0.75$ which is roughly doubled with respect to the corresponding Newtonian case. Another interesting feature emerging from the second row of images of Fig.~\ref{fig:6} is that the formation of multiple neckings is significantly suppressed by viscoelasticity. In particular, the droplet still breaks-up in three droplets, but their size is different, with the central droplet being much smaller that the other two. Being interested in using a uniform and confined shear flow to generate quasi monodisperse emulsions by controlled break-up~\cite{Sibillo06,Renardy07}, Fig.~\ref{fig:6} suggests to use caution in presence of non-Newtonian phases. 


\begin{figure}[pth!]
\subfigure[\,\,De = 0]
    {
        \includegraphics[scale=0.56]{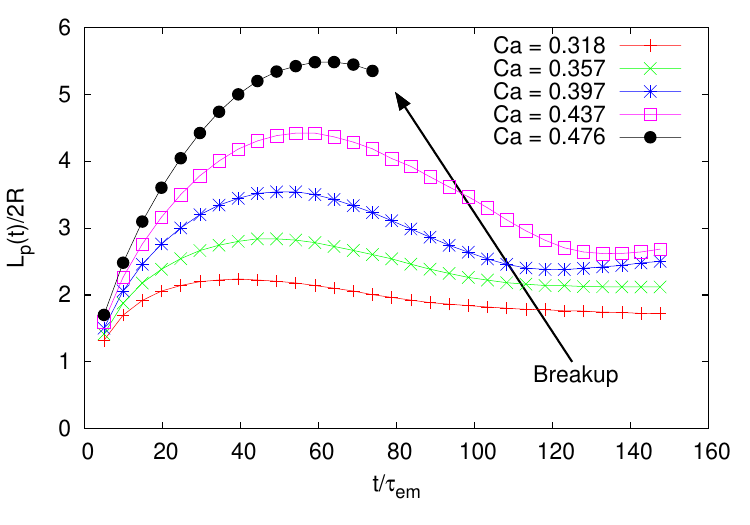}
    }    \\
\subfigure[\,\,De = 0.1, $L^2=10^2$]
    {
        \includegraphics[scale=0.56]{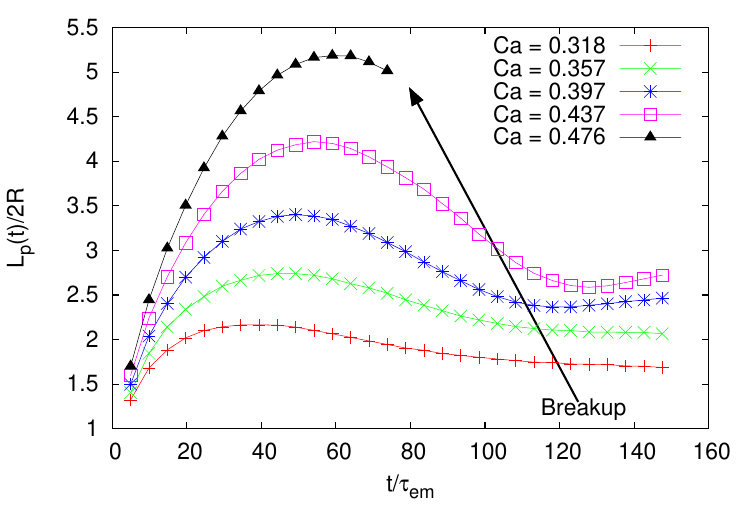}
    }    \\
\subfigure[\,\,De = 2.0, $L^2=10^2$]
    {
        \includegraphics[scale=0.56]{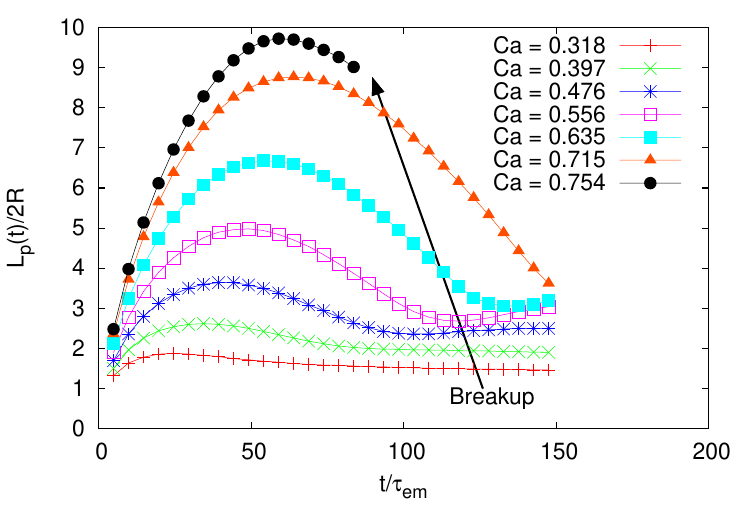}
    }\\
\caption{Evolution of the dimensionless droplet length after the startup of a shear flow for various Capillary numbers and Deborah numbers for a fixed confinement ratio $2R/H = 0.78$ and finite extensibility parameter $L^2=10^2$. Since the shape of highly deformed droplets may deviate from an ellipsoid, we estimated the droplet elongation from the projection of the droplet length ($L_p$) in the velocity direction. The viscous ratio between the droplet phase and the matrix phase is kept fixed to $\lambda=\eta_D/\eta_M=1$, independently of the degree of viscoelasticity. Similarly to Figs.~\ref{fig:3}-\ref{fig:6}, we use the emulsion time $\tau_{\mbox{\tiny{em}}}$ (see Eq.~\eqref{emulsiontime}) as a unit of time. \label{fig:7}}
\end{figure}

Thus, the effects of viscoelasticity on the critical Capillary number appear more sizeable in the case with a larger confinement ratio. This is complemented by the results reported in Fig.~\ref{fig:7}, where we show the dimensionless droplet elongation $L_p/2R$ as a function of time for several values of $\mbox{De}$ and $\mbox{Ca}$. Since the shape of highly deformed and confined droplets deviates from an ellipsoid, we estimated the droplet elongation from the projection of the droplet length ($L_p$) in the velocity direction.  In Panel (a) of Fig.~\ref{fig:7} we show the results for a Newtonian droplet with Capillary number ranging in the interval $0.318 \le \mbox{Ca} \le 0.476$, with the critical Capillary number being equal to $\mbox{Ca}_{\mbox{\tiny{cr}}} = 0.476$ (see Fig.~\ref{fig:5}). Before break-up, the droplet elongation reaches a maximum value and then it breaks while retracting, which is another signature of the triple break-up discussed before. The maximum elongation increases with the Deborah number (Panel (b)-(c) of Fig.~\ref{fig:7}) and the increase of the maximum elongation goes together with an increase of the critical Capillary number: the maximum elongation is indeed doubled when moving from $\mbox{De}=0.1$ to $\mbox{De}=2.0$.


\begin{figure}[pth!]
\subfigure[\,\,]
{
\includegraphics[scale=0.5]{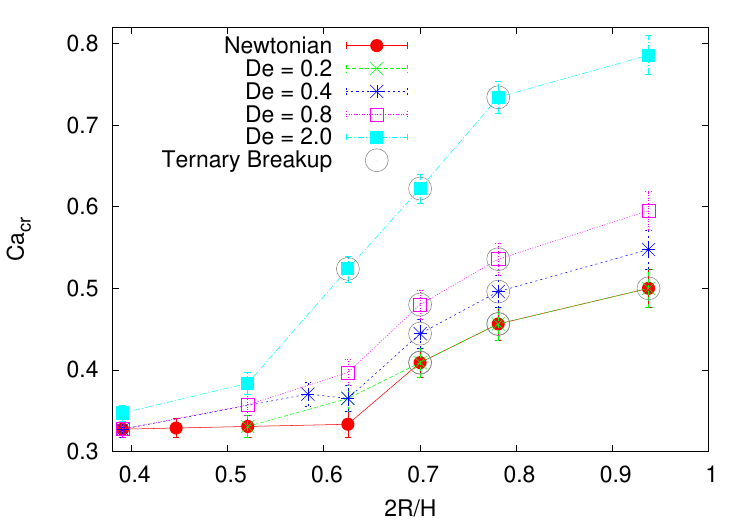}
}\\
\subfigure[\,\,]
{
\includegraphics[scale=0.5]{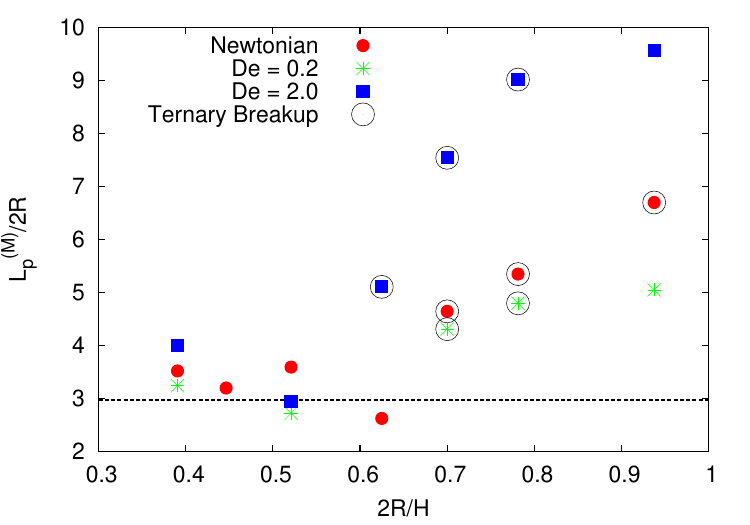}
}\\
\subfigure[\,\,]
{
\includegraphics[scale=0.5]{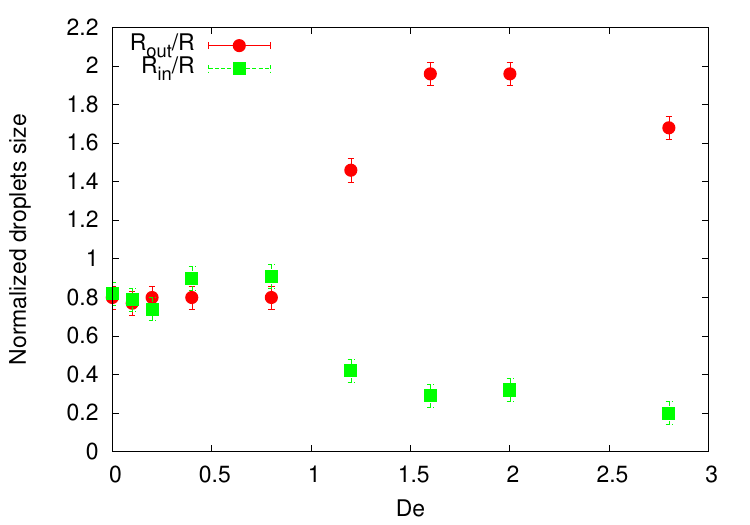}
}\\
\caption{Panel (a): Critical Capillary number for break-up as a function of confinement ratio for systems with finite extensibility parameter $L^2 = 10^2$. The viscous ratio between the polymeric droplet phase and the matrix phase is kept fixed to $\lambda=\eta_D/\eta_M=1$. Different Deborah numbers are considered, by letting the polymer relaxation time $\tau_P$ in Eq.~\eqref{FENE} changing in the interval $0 \le \tau_P \le 7000$ lbu. Black open circles indicate situations where multiple neckings occur. Panel (b): data analyzed in Panel (a) are reported in terms of the dimensionless maximum elongation of the droplet $L^{(M)}_p/2R$. Panel (c): we plot the dimensionless size of the outer ($R_{out}/R$) and inner ($R_{in}/R$) daughter droplets in the triple break-up (see also Fig.~\ref{fig:6}). Up to $\mbox{De} \approx 1$, droplets break into roughly equal sized daughter droplets, but for $\mbox{De}>1$ there is substantial change in the size of these daughter droplets, as the size of inner (outer) daughter droplet starts decreasing (increasing) rapidly \label{fig:8}}
\end{figure}

Overall, there are two main messages conveyed by Figs.~\ref{fig:3}-\ref{fig:7}. First, it is evident that viscoelasticity has a stabilizing effect on droplet break-up~\cite{Flumerfelt72,Elmendorp,Mighri,Lerdwijitjarud03,Lerdwijitjarud04,AggarwalSarkar07,AggarwalSarkar08,GuidoRev,Verhulst09a,Verhulst09b}, with a larger effect in presence of a larger confinement ratio~\cite{Cardinaels11}. Second, the formation of multiple neckings - a distinctive feature of break-up of confined droplets - is also affected by the presence of viscoelasticity. These statements are better complemented by the results reported in Fig~\ref{fig:8}, which give an overview of all the various numerical simulations performed, at changing confinement ratio and degree of viscoelasticity, while keeping the finite extensibility of the polymers fixed to $L^2=10^2$. In Panel (a) of Fig.~\ref{fig:8}, we report the critical Capillary number $\mbox{Ca}_{\mbox{\tiny{cr}}}$ as a function of the confinement ratio. For Newtonian droplets, the role of confinement is almost insignificant up to $2R/H = 0.625$, whereas for larger confinement ratio a monotonous increase of $\mbox{Ca}_{\mbox{\tiny{cr}}}$ is observed. The emergence of this up-turn in $\mbox{Ca}_{\mbox{\tiny{cr}}}$ is a direct consequence of the change of the break-up mechanism. Up to $\mbox{De} \approx 1$, $\mbox{Ca}_{\mbox{\tiny{cr}}}$ only slightly increases upon increasing $\mbox{De}$. When $\mbox{De}>1$, however, the change in $\mbox{Ca}_{\mbox{\tiny{cr}}}$ significantly increases with $2R/H$. The black open circles indicate situations where ternary break-up is observed.  We notice that the addition of polymers to droplets for the highest confinement ratio considered ($2R/H = 0.93$) is enough to remove ternary break-up, independently of the degree of viscoelasticity. In addition to the critical Capillary number, in Panel (b) of Fig.~\ref{fig:8}, we report the maximum dimensionless elongation of the droplet, $L^{(M)}_p/2R$, as a function of confinement ratio. It is clear that the trends for $\mbox{Ca}_{\mbox{\tiny{cr}}}$ and $L^{(M)}_p/2R$ are quite similar. Indeed, $L^{(M)}_p/2R$ starts to increase at approximately the same degree of confinement where $\mbox{Ca}_{\mbox{\tiny{cr}}}$ shows the up-turn (see Panel (a) of Fig.~\ref{fig:8}). We have also drawn a horizontal dashed line to show the cutoff ($L_p/R =	5.95$) predicted by Janssen {\it et al.}~\cite{Janssen10} above which the Rayleigh-Plateau instability sets-in. As shown in Figs.~\ref{fig:6} and \ref{fig:8} , viscoelasticity has an effect on the triple break-up of confined droplets. To better quantify this effect, we measured the size of the ``outer'' and ``inner'' daughter droplets. The dimensionless sizes of such daughter droplets, $R_{\mbox{\tiny{out}}}/R$ and $R_{\mbox{\tiny{in}}}/R$, are shown as a function of $\mbox{De}$ in Panel (c) of Fig.~\ref{fig:8}. It is clear from this plot that up to $\mbox{De} \approx 1$ droplets break into roughly equal sized daughter droplets, but for $\mbox{De}>1$ there is substantial change, as the size of inner (outer) daughter droplet starts decreasing (increasing) rapidly by increasing $\mbox{De}$.

\begin{figure}[pth!]
\subfigure[\,\,De = 2.0, 2R/H = 0.78, $L^2 = 10^2$]
    {
        \includegraphics[scale=0.5]{plot_L_Lsqr_100_taup_5000}
    }\\
\subfigure[\,\,De = 2.0, 2R/H = 0.78, $L^2 = 10^4$]
    {
        \includegraphics[scale=0.5]{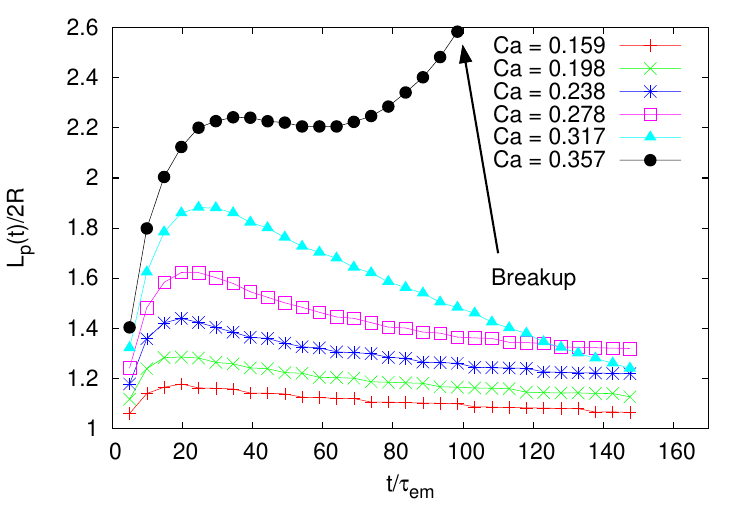}
    }  \\  
\subfigure[\,\,2R/H = 0.78]
    {
        \includegraphics[scale=0.5]{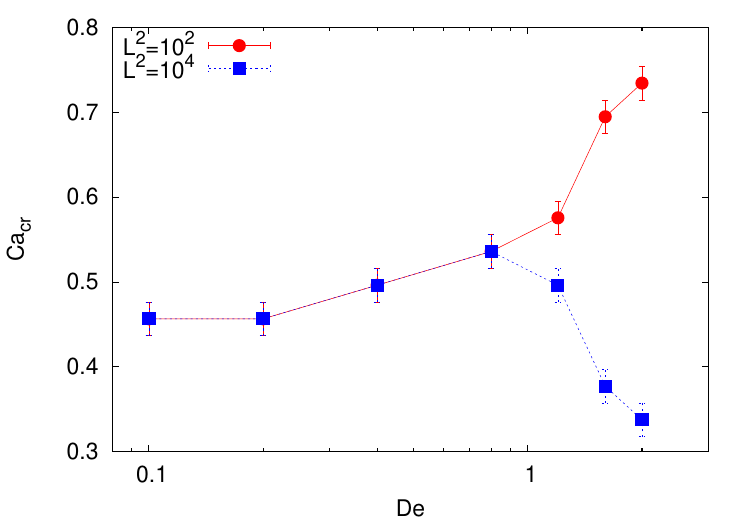}
    }    \\
\caption{Panel (a): evolution of the dimensionless droplet length after the startup of a shear flow for various Capillary numbers. We fix both the Deborah number ($\mbox{De}=2.0$), the confinement ratio ($2R/H = 0.78$), and the finite extensibility parameter $L^2=10^2$ (data already shown in Panel (c) of Fig.~\ref{fig:7}). Panel (b): same as Panel (a) with an increased finite extensibility parameter, $L^2=10^4$. The increase of $L^2$ determines different elongational properties and is affecting the critical Capillary number. Panel (c): the critical Capillary number is reported as a function of the Deborah number for the two values of $L^2$ considered in Panels (a)-(b). In all cases, the viscous ratio between the polymeric droplet phase and the matrix phase is kept fixed to $\lambda=\eta_D/\eta_M=1$.\label{fig:9}}
\end{figure}

\begin{figure}[pth!]
\subfigure[\,\,t/$\tau_{\mbox{\tiny{em}}}$=25, 2R/H = 0.78, Ca=0.75, De=2.0, L$^2$=10$^2$]
    {
        \includegraphics[scale=0.038]{confined_stronglyvisco_t_25000_up}
    }    
\subfigure[\,\,t/$\tau_{\mbox{\tiny{em}}}$=75, 2R/H = 0.78, Ca=0.75, De=2.0, L$^2$=10$^2$]
    {
        \includegraphics[scale=0.038]{confined_stronglyvisco_t_100000_up}
    }    
\subfigure[\,\,t/$\tau_{\mbox{\tiny{em}}}$=100, 2R/H = 0.78, Ca=0.75, De=2.0, L$^2$=10$^2$]
    {
        \includegraphics[scale=0.038]{confined_stronglyvisco_t_125000_up}
    }
\\
\subfigure[\,\,t/$\tau_{\mbox{\tiny{em}}}$=25, 2R/H = 0.78, Ca=0.35, De=2.0, L$^2$=10$^4$]
    {
        \includegraphics[scale=0.038]{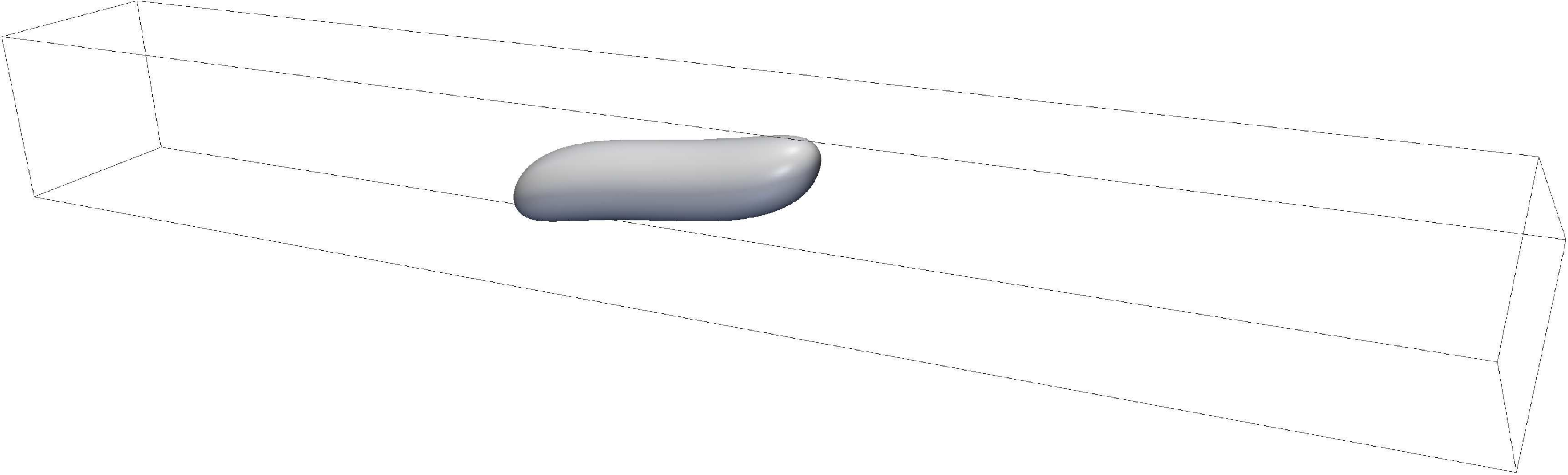}
    }    
\subfigure[\,\,t/$\tau_{\mbox{\tiny{em}}}$=75, 2R/H = 0.78, Ca=0.35, De=2.0, L$^2$=10$^4$]
    {
        \includegraphics[scale=0.038]{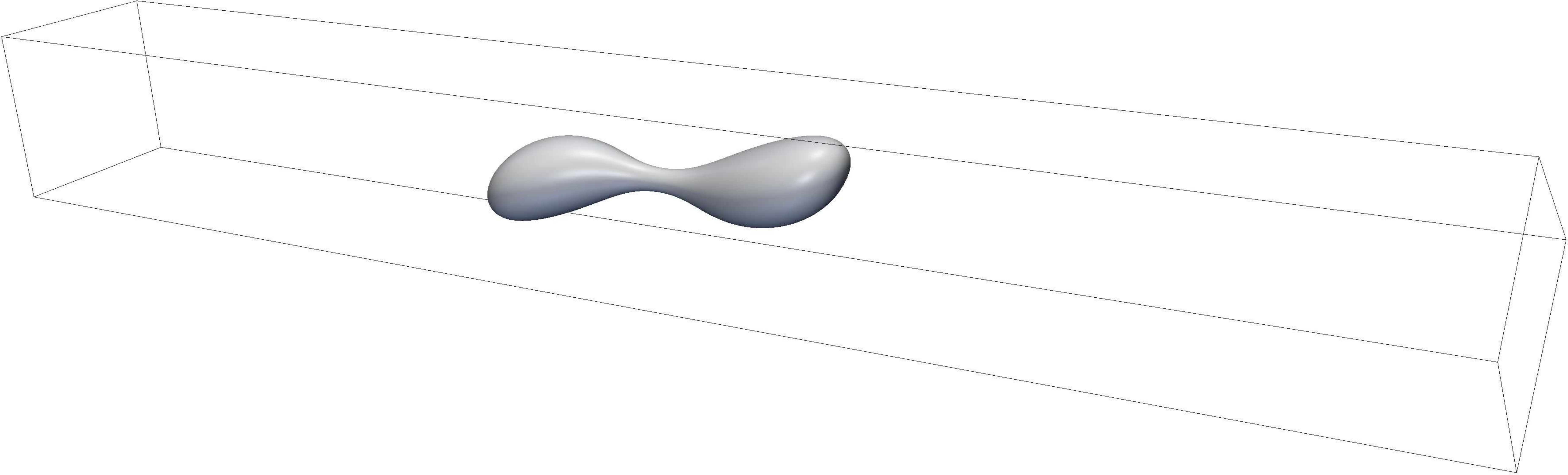}
    }    
\subfigure[\,\,t/$\tau_{\mbox{\tiny{em}}}$=100, 2R/H = 0.78, Ca=0.35, De=2.0, L$^2$=10$^4$]
    {
        \includegraphics[scale=0.038]{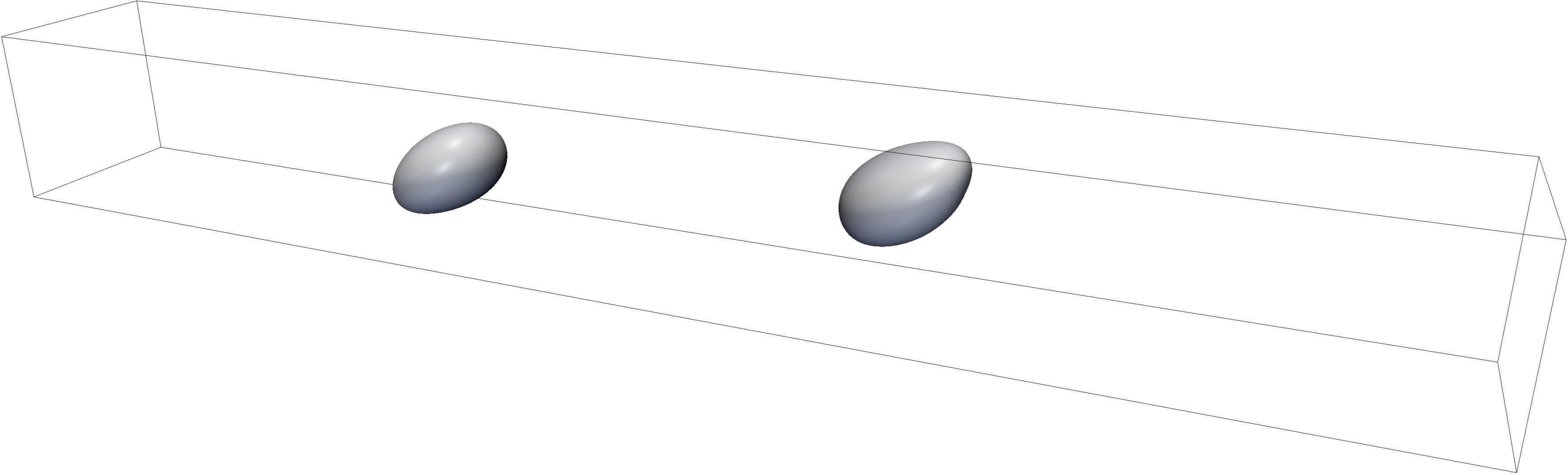}
    }
\caption{Influence of the finite extensibility parameter $L^2$ in the break-up after the startup of a shear flow with confinement ratio $2R/H = 0.78$ and fixed Deborah number $\mbox{De}=2.0$. We report the time history of droplet deformation and break-up including 3 representative snapshots: initial deformation (left column, $t = 25 \tau_{\mbox{\tiny{em}}}$); deformation prior to break-up (middle column, $t = 75 \tau_{\mbox{\tiny{em}}}$); post-break-up frame (right column, $t = 100\tau_{\mbox{\tiny{em}}}$). We use the emulsion time $\tau_{\mbox{\tiny{em}}}$ (see Eq.~\eqref{emulsiontime}) as a unit of time. In all cases, the viscous ratio between the droplet phase and the matrix phase is kept fixed to $\lambda=\eta_D/\eta_M=1$, independently of the degree of viscoelasticity.\label{fig:10}}
\end{figure}

So far, we have kept the finite extensibility $L$ of the polymers fixed. However, as $L$ increases, the polymer dumbbell becomes more extensible and the maximum level of stress attainable is increased. More quantitatively, in a homogeneous steady uniaxial extension, the extensional viscosity of the polymers increases proportionally to $L^2$ and it becomes infinite in the limit $L^2 \gg 1 $~\cite{bird87,Lindner03}. It is also noted that a simple shear flow can always be decomposed into two parts: an antisymmetric one which provides a rigid-like clockwise rotation of the droplet, and a symmetric part corresponding to an elongational flow, which tends to elongate and orientate the droplet along $\theta=\pi/4$~\cite{Rioual}. Thus, at changing the elongational viscosity of the droplet, we expect a different response under shear flow. A further hint that the elongational properties of the droplet are affecting droplet deformation and subsequent break-up is provided by Fig.~\ref{fig:9}, where we report the dimensionless droplet elongation $L_p/2R$ as a function of time for several values of $\mbox{Ca}$ and fixed $\mbox{De}=2.0$. Two different values of $L^2$ are considered: $L^2=10^2$ (Panel (a), data already shown in Fig.~\ref{fig:7}) and $L^2=10^4$ (Panel (b)). The maximum elongation of the droplet is inhibited by changing the maximum elongation of the polymers and break-up takes place at a much smaller Capillary number, $\mbox{Ca}_{\mbox{\tiny{cr}}}\approx 0.34$. Also, while in the case with $L^2=10^2$ the droplet first retracts and then breaks with a triple break-up (see also Fig.~\ref{fig:6}), this does not seem to be the case for $L^2=10^4$. Panel (c) of Fig.~\ref{fig:9} reports $\mbox{Ca}_{\mbox{\tiny{cr}}}$ as a function of the Deborah number for the two values of $L^2$ considered. Up to $\mbox{De}=1$, the behaviour of the critical Capillary number is essentially the same, witnessing an irrelevant role of viscoelasticity. However, for a Deborah above unity, an opposite effect is found: while for $L^2=10^2$ the critical Capillary number is increasing with the Deborah number, at much larger $L^2$ the critical Capillary number decreases.  The reason for this bifurcation is found in a different mechanism of break-up, as evidenced by Fig.~\ref{fig:10}, where we report 3D snapshots showing deformation and subsequent break-up for the droplets with both $L^2=10^2$ and $L^2=10^4$ in post-critical situations. For $L^2 = 10^2$ (panels (a)-(c)), the droplet first elongates above the critical elongation where the Rayleigh-Plateau instability develops and then breaks during retraction (data already shown in Fig.~\ref{fig:9}). For $L^2=10^4$ (panels (d)-(f)) the droplet does not elongate, it just deforms and breaks very similarly to the unconfined case (Panels (a)-(c) of Fig.~\ref{fig:3}). Consequently, the critical Capillary number is decreased to $\mbox{Ca}_{\mbox{\tiny{cr}}} \approx 0.34$, thus becoming much more comparable with the unconfined value (see Fig.~\ref{fig:3}). For completeness, we repeated the numerical simulations in the unconfined case and we could not estimate a significant difference in the critical Capillary number $\mbox{Ca}_{\mbox{\tiny{cr}}}$ at changing the finite extensibility parameter $L^2$. This lends further support to the idea that a non trivial interplay between confinement and viscoelasticity is at the core of the observed behaviour for the critical Capillary number.


\section{Force Balance inside the droplet}\label{sec:forcebalance}

\begin{figure}[pth!]
{
\includegraphics[scale=0.07]{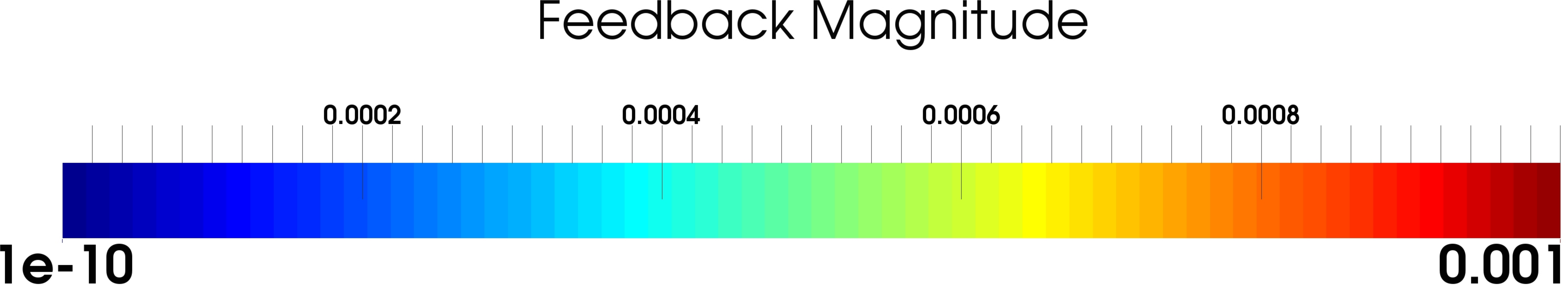}
}
\\
\subfigure[\,\,t/$\tau_{\mbox{\tiny{em}}}$=25, 2R/H = 0.4, Ca=0.33, De=2.0]
{
\includegraphics[scale=0.04]{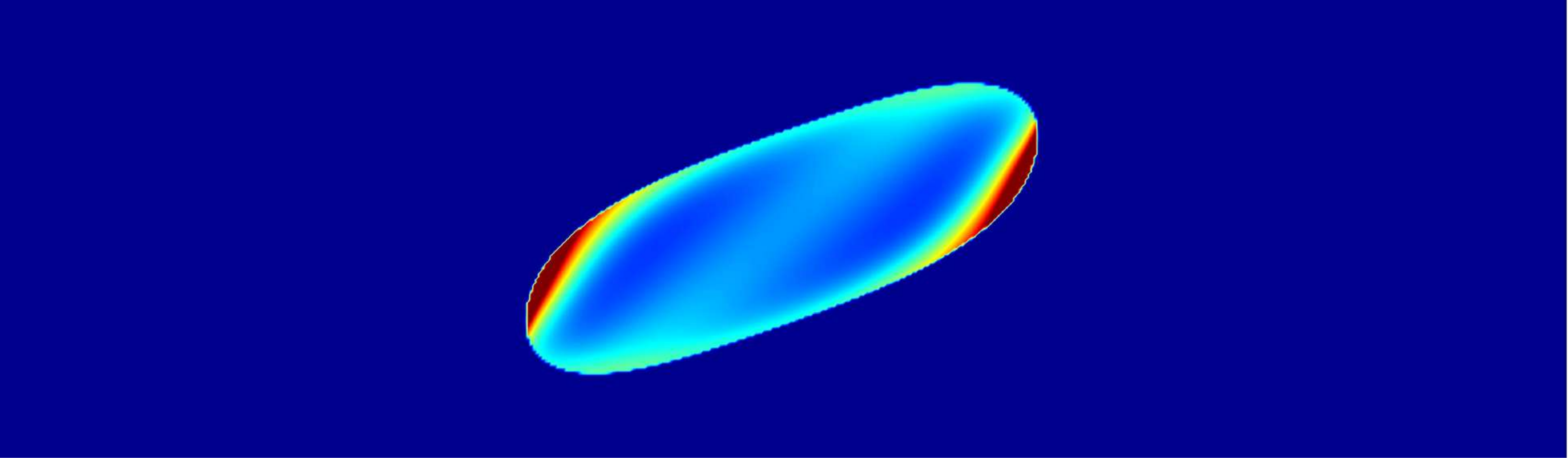}
}    
\subfigure[\,\,t/$\tau_{\mbox{\tiny{em}}}$=100, 2R/H = 0.4, Ca=0.33, De=2.0]
{
\includegraphics[scale=0.04]{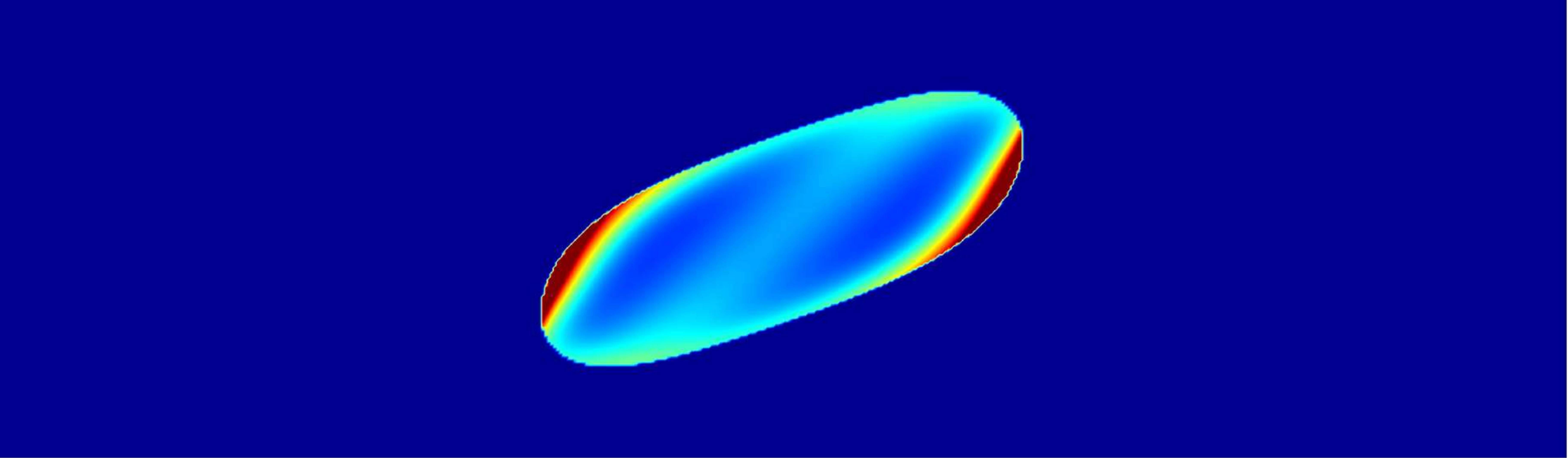}
}    
\subfigure[\,\,t/$\tau_{\mbox{\tiny{em}}}$=125, 2R/H = 0.4, Ca=0.33, De=2.0]
{
\includegraphics[scale=0.04]{unbounded_stress_stronglyvisco_75000}
}
\\
\subfigure[\,\,t/$\tau_{\mbox{\tiny{em}}}$=25, 2R/H = 0.78, Ca=0.47, De=2.0]
{
\includegraphics[scale=0.036]{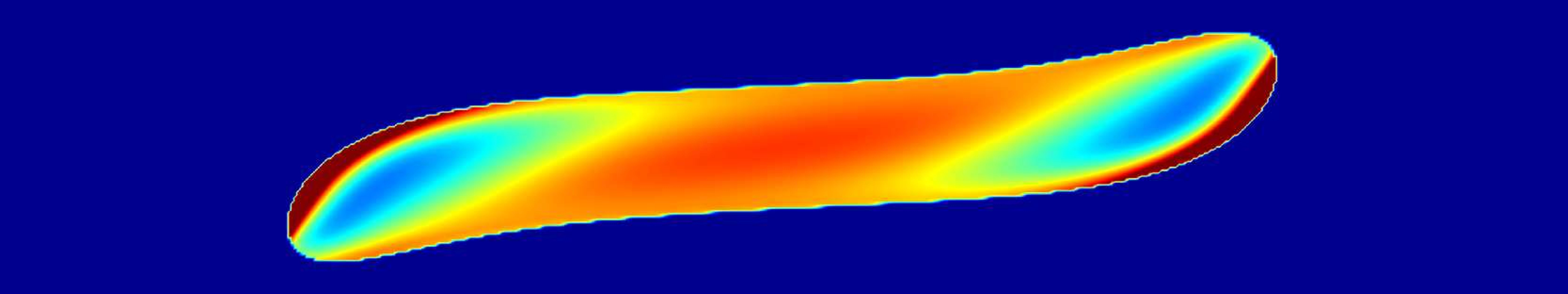}
}    
\subfigure[\,\,t/$\tau_{\mbox{\tiny{em}}}$=100, 2R/H = 0.78, Ca=0.47, De=2.0]
{
\includegraphics[scale=0.036]{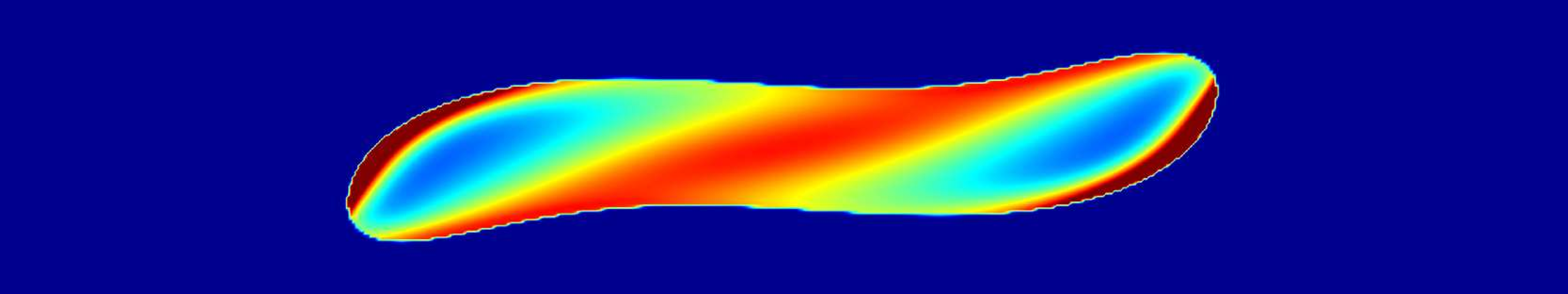}
}    
\subfigure[\,\,t/$\tau_{\mbox{\tiny{em}}}$=125, 2R/H = 0.78, Ca=0.47, De=2.0]
{
\includegraphics[scale=0.036]{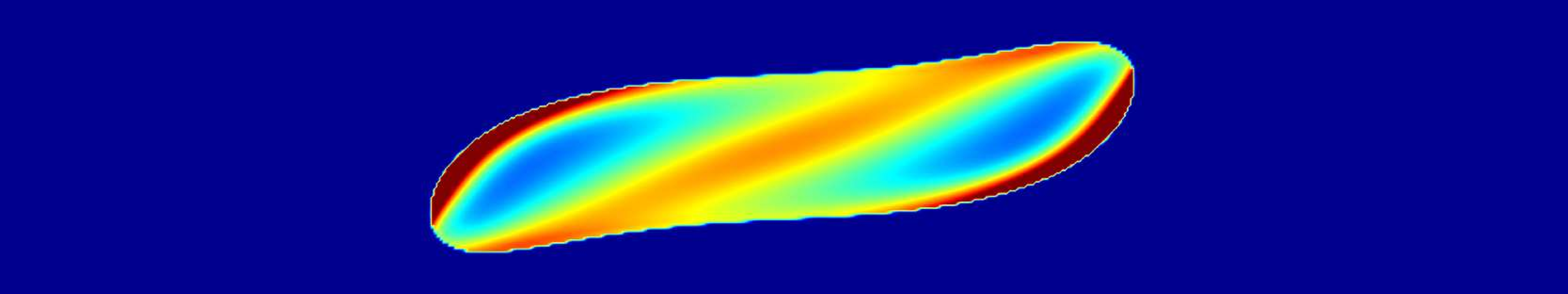}
}
\caption{Feedback stress magnitude in the shear plane ($xz$ plane at $y=L_y/2$) for the viscoelastic data with $\mbox{De}=2.0$ reported in Figs.~ \ref{fig:3} and \ref{fig:5}. We use the emulsion time $\tau_{\mbox{\tiny{em}}}$ (see Eq.~\eqref{emulsiontime}) as a unit of time. In both confinement ratios, the Capillary number is such that it corresponds to the post-critical condition for the corresponding Newtonian case ($\mbox{De}=0$). Gradients in the polymeric stress are modulated in space and more pronounced in the confined case, which qualitatively explains the larger increase in the Capillary number at break-up. \label{fig:11}}.
\end{figure}

In Sec.~\ref{sec:dropletbreakup} we have analyzed the behaviour of confined droplets under shear flows and determined the associated critical Capillary number. Our simulations have provided easy access to quantities such as droplet deformation and orientation showing a non trivial interplay between confinement and viscoelasticity. Indeed, by increasing the confinement ratio, we have seen that two opposite behaviours can take place, dependently on the finite extensibility parameter of the polymers. Simulations also allow to monitor the velocity flow field, pressure field and polymers feedback stress inside the droplet. The goal of the present section is therefore to complement the results discussed in Sec.~\ref{sec:dropletbreakup} by directly monitoring the various forces contributions which are present in Eqs~\eqref{NS} and \eqref{FENE}.\\
To start, in Fig.~\ref{fig:11} we show some snapshots of the feedback stress in the shear plane ($xz$ plane at $y=L_y/2$) for a Deborah number above unity ($\mbox{De} = 2.0$) and fixed $L^2=10^2$. Data are the same reported in panels (g)-(i) of Figs.~\ref{fig:3} and \ref{fig:5}: the top and bottom rows correspond to the confinement ratios $2R/H = 0.4$ and $2R/H = 0.78$, respectively. We see that the maximum of the feedback stress is slightly above the tip of the droplet at the back, and slightly below the tip of the droplet at the front~\cite{Verhulst09a}. Also, the spatial modulation is suggesting that the polymer feedback stress is providing a resistance against elongation in the direction $\theta=\pi/4$ with respect to the flow direction, which echoes the discussion on the elongational viscosity done in the previous section. 


\begin{figure}[pth!]
\subfigure[\,\,2R/H = 0.4, Ca=0.32, De=0]
    {
        \includegraphics[scale=0.5]{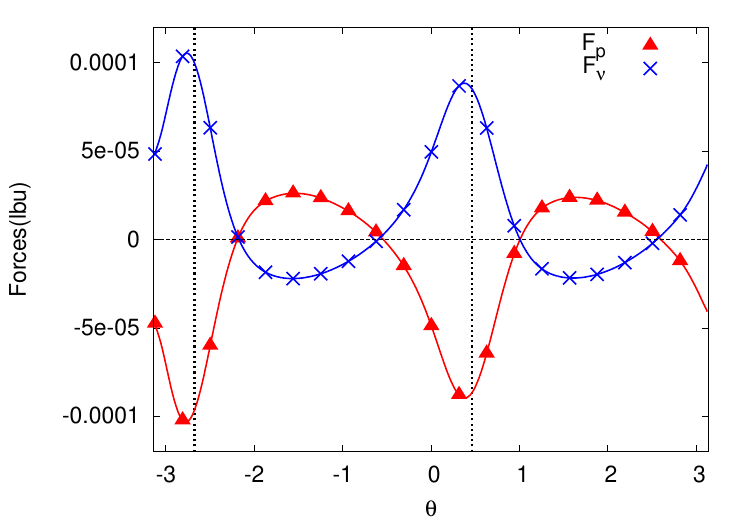}
    }    
\subfigure[\,\,2R/H = 0.78, Ca=0.32, De=0]
    {
        \includegraphics[scale=0.5]{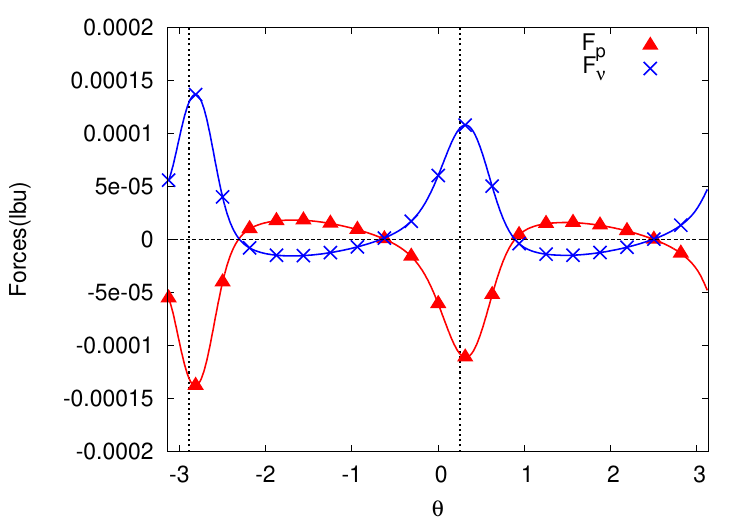}
    }    
\\
\subfigure[\,\,2R/H = 0.4, Ca=0.34, De=2.0, $L^2 = 10^2$]
    {
        \includegraphics[scale=0.5]{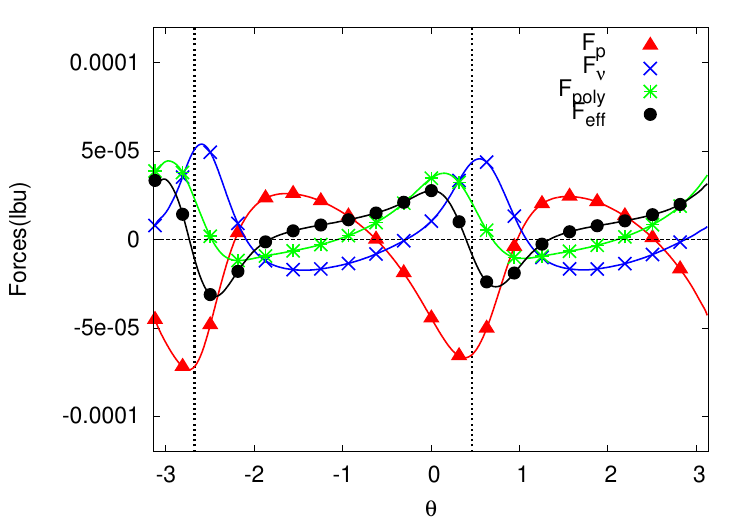}
    }    
\subfigure[\,\,2R/H = 0.78, Ca=0.32, De=2.0, $L^2 = 10^2$]
    {
        \includegraphics[scale=0.5]{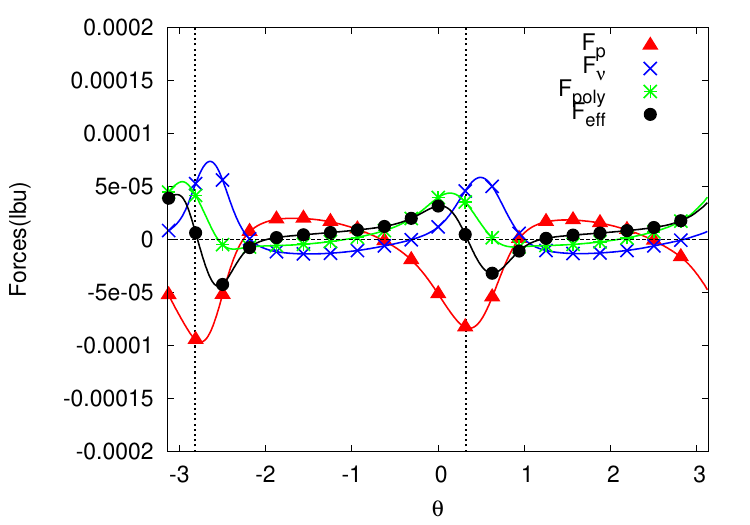}
    }  
\\
\caption{We report the forces contributions resulting from Eq.~\eqref{NS} in the shear plane ($xz$ plane at $y=L_y/2$). The Capillary number is fixed, $\mbox{Ca} = 0.32$, corresponding to steady-states for all the cases studied. We project the viscous forces (${\bm F}_{\nu}$), the pressure forces (${\bm F}_{p}$), and the viscoelastic forces (${\bm F}_{poly}$, where applicable) of Eq.~\eqref{NS} in the radial direction at a given distance ($R/10$ lbu) from the interface. The force balance is then studied as a function of the angular position $\theta$ (see Fig~\ref{fig:1}) from the flow direction. Left panel figures are related to a confinement ratio $2R/H = 0.4$, whereas the right panel ones refer to $2R/H = 0.78$. Different Deborah numbers are considered. Panels (a)-(b) show the force balance for the Newtonian case ($\mbox{De}=0$); Panels (c)-(d) show the viscoelastic case with $\mbox{De} = 2.0$ and $L^2 = 10^2$.  The vertical dashed lines show the orientation angle of the droplet, computed as the one of the equivalent ellipsoid. All forces are reported in lbu (LBM units). \label{fig:12}}
\end{figure}

\begin{figure}[pth!]
\subfigure[\,\,2R/H = 0.78, Ca = 0.32, De=2.0,  $L^2=10^2$]
    {
        \includegraphics[scale=0.5]{rw_08_plot_radial_Lsqr_100_200}
    }    
\subfigure[\,\,2R/H = 0.78, Ca = 0.32, De=2.0, $L^2=10^3$]
    {
        \includegraphics[scale=0.5]{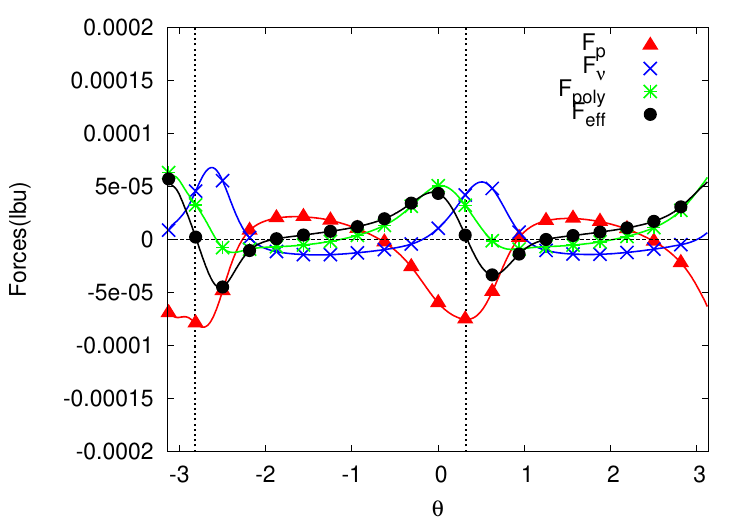}
    }
\\
\subfigure[\,\,2R/H = 0.78, Ca = 0.32, De=2.0, $L^2=5 \times 10^3$]
    {
        \includegraphics[scale=0.5]{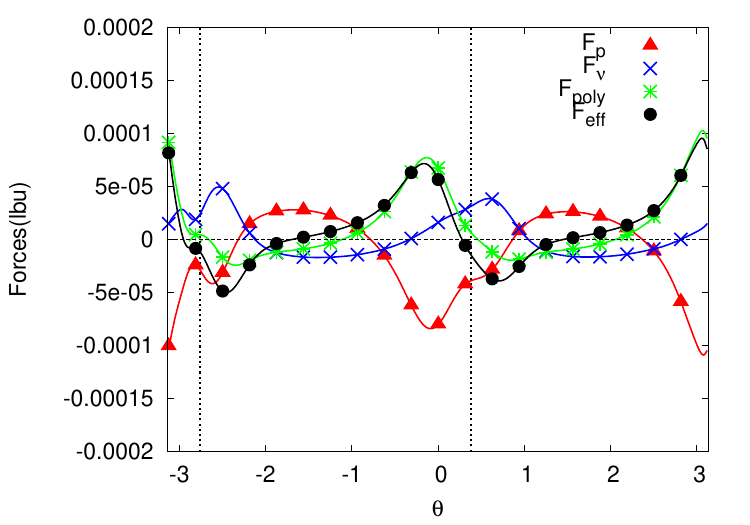}
    }    
\subfigure[\,\,2R/H = 0.78, Ca = 0.32, De=2.0, $L^2=10^4$]
    {
        \includegraphics[scale=0.5]{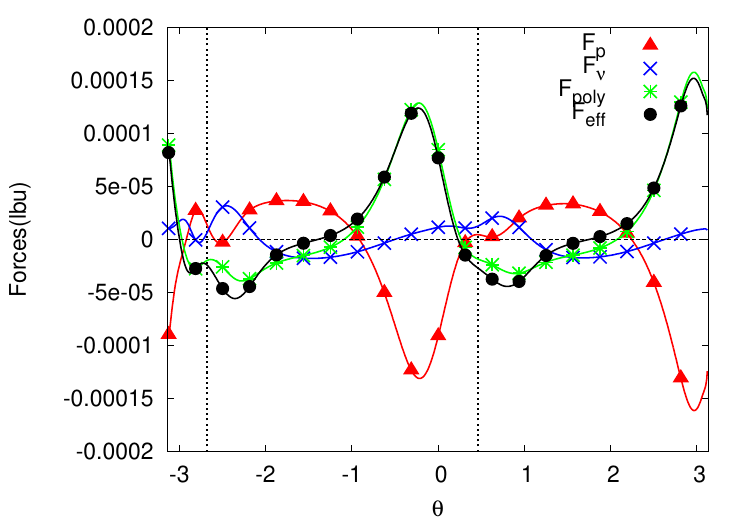}
    }
\\
\caption{We repeat the analysis of Fig.~\ref{fig:12} for different values of the finite extensibility parameter $L^2$, by keeping the Deborah number fixed to $\mbox{De}=2.0$ and the Capillary number fixed to $\mbox{Ca}=0.32$. The confinement ratio is kept fixed to $2R/H = 0.78$. All forces are reported in lbu (LBM units). \label{fig:13}}
\end{figure}

To quantitatively understand both the role of confinement and viscoelasticity, in Fig~\ref{fig:12} we show the forces contributions at the stationary state for a droplet with two confinement ratios, $2R/H = 0.4$ and $2R/H = 0.78$, and finite extensibility parameters, $L^2=10^2$. We quantitatively compare the Newtonian ($\mbox{De}=0$) and the viscoelastic case with Deborah number above unity ($\mbox{De}=2.0$). Data are shown for the same Capillary number $\mbox{Ca}=0.32$, corresponding to pre-critical conditions for the Newtonian ($\mbox{De}=0$) droplet in the smaller ($2R/H = 0.4$) confinement ratio analyzed. Working in the shear plane ($xz$ plane at $y=L_y/2$), we project the viscous forces (${\bm F}_{\nu}={\bm \nabla} \cdot \left(\eta_{A} ({\bm \nabla} {\bm u}+({\bm \nabla} {\bm u})^{T})\right)$), the pressure forces (${\bm F}_{p}=- {\bm \nabla}P$), and the viscoelastic forces (${\bm F}_{\mbox{\tiny{poly}}}=\frac{\eta_P}{\tau_P}{\bm \nabla} \cdot [f(r_P){\bm {\bm {\mathcal C}}}]$, where applicable) of Eq.~\eqref{NS} in the radial direction at a given distance ($R/10$ lbu) from the interface. The force balance is then studied as a function of the angular position $\theta$ (see Fig~\ref{fig:1}) from the flow direction. In the Newtonian case (panels (a)-(b)) the pressure forces are well balancing with the viscous forces and the structure of the angular modulation of the forces is quite similar in the two confinement ratios analyzed. The negative radial peak of the pressure forces is located in correspondence of the major semi-axis of the droplet (indicated with a dotted line) where the curvature is larger. We remark, however, that elongated droplets are stabilized by confinement, and therefore break at a larger Capillary number with a triple break-up. The structure of the force balance is changed by the introduction of the viscoelastic stresses (Panels (c)-(d)). In order to properly analyze these figures, one has to remark that viscoelastic forces provide a contribution to the shear forces. This happens in simple shear flows and also for weak viscoelasticity~\cite{bird87,Herrchen97}, where we expect that the viscoelastic stresses closely follow the viscous stresses, i.e. $\frac{\eta_P}{\tau_P}{\bm \nabla} \cdot [f(r_P){\bm {\bm {\mathcal C}}}] \approx {\bm \nabla} \cdot \left(\eta_{P} ({\bm \nabla} {\bm u}+({\bm \nabla} {\bm u})^{T})\right)$. Obviously, this cannot be the case when viscoelasticity is enhanced and the Deborah number is above unity. For this reason, to better visualize the importance of the viscoelastic forces in comparison with the Newtonian case, we have defined the {\it effective} force (${\bm F}_{\mbox{\tiny{eff}}}$) as 
\be\label{eq:effectiveforce}
{\bm F}_{\mbox{\tiny{eff}}}=\frac{\eta_P}{\tau_P}{\bm \nabla} \cdot [f(r_P){\bm {\bm {\mathcal C}}}]-{\bm \nabla} \left(\eta_{P} ({\bm \nabla} {\bm u}+({\bm \nabla} {\bm u})^{T})\right).
\ee
Since all our simulations are performed with the same shear viscosity inside the droplet, the effective force gives us an idea of how much the viscoelastic system differs from the corresponding Newtonian system with the same viscosity. If present (${\bm F}_{\mbox{\tiny{eff}}} \neq 0$), this change is solely attributed to viscoelasticity. The effective force for the cases with $L^2=10^2$ (panels (c)-(d)) is peaked in correspondence of the droplet semi-axes, with a negative (positive) radial contribution along the major (minor) semi-axis. This supports the discussion done in the previous section, in that the viscoelastic forces provide a resistance against elongation in the direction $\theta=\pi/4$, although the peaks appear in correspondence of slightly different angles than $\pi/4$ as the droplet is already deformed and deviates from an ellipsoidal shape, especially in the larger confinement ratio. To quantify the role of the finite extensibility parameter $L^2$ in the force balance, we repeated the analysis shown in Fig.~\ref{fig:12} for a fixed Capillary number ($\mbox{Ca}=0.32$), fixed Deborah number ($\mbox{De}=2.0$), fixed confinement ratio ($2R/H = 0.78$), and for different values of the finite extensibility parameter ranging in the interval $10^2 \le L^2 \le 10^4$. Results are reported in Fig.~\ref{fig:13}. It is clear that as $L^2$ increases, polymer forces develop along the orientation axes of the droplet, preventing the droplet from being elongated. In particular, a net positive radial contribution along the minor semi-axis starts growing at $L^2=10^3$ with increasing magnitude at increasing $L^2$. 


\section{Conclusions}\label{sec:conclusions}

The deformation and break-up of Newtonian/viscoelastic droplets in systems with a Newtonian matrix have been studied in confined shear flow. We have proposed numerical simulations based on a hybrid algorithm combining lattice-Boltzmann models (LBM) and finite differences schemes, the former used to model the Navier-Stokes equations, and the latter used to model the kinetics of polymers using the constitutive equations for finitely extensible non-linear elastic dumbbells with Peterlin's closure (FENE-P). Simulations provide easy access to quantities such as droplet deformation and orientation as well as the velocity flow field, viscous and viscoelastic stresses, and pressure field. Various messages are conveyed by our analysis. It is evident that droplet viscoelasticity has a stabilizing effect on droplet break-up~\cite{Flumerfelt72,Elmendorp,Mighri,Lerdwijitjarud03,Lerdwijitjarud04,AggarwalSarkar07,AggarwalSarkar08,GuidoRev,Verhulst09a,Verhulst09b}. The effect is larger in presence of a larger confinement ratio. In particular, in agreement with some recent experiments~\cite{Cardinaels11}, we have found that the formation of multiple neckings, which is acknowledged as a distinctive feature of confined break-up~\cite{Sibillo06,Janssen10}, is also affected by the presence of viscoelasticity: it visibly changes as soon as the ratio of fluid relaxation time to droplet emulsion time (i.e. the Deborah number) becomes of the order of 1. A non trivial interplay between confinement and the maximum elongation of the polymers has also emerged. With the use of numerical simulations we had the opportunity to change separately the viscous ratio in the Newtonian phases, the maximum extension of the polymers, and the degree of viscoelasticity, thus allowing for a systematic analysis of the viscoelastic effects while keeping the shear viscosity of the droplet fixed to the reference Newtonian case. In particular, by increasing the finite extensibility of the polymers,  it is observed that the resistance against elongation may be enough to prevent both droplet elongation and subsequent triple break-up, thus altering significantly the critical Capillary number for viscoelastic droplets under confinement.\\
For future investigations, it is surely warranted a complementary study to highlight the role of matrix viscoelasticity on the break-up properties of confined droplets. Also, as an upgrade of complexity, it would be extremely interesting to study other more structured flows in confined geometries, like flow-focusing devices with viscoelastic phases~\cite{Arratia,Garstecky}. Complementing the experimental results with the help of numerical simulations would be of extreme interest. Simulations can indeed be used to perform in-silico  comparative studies, at changing the model parameters, to shed lights on the complex properties of viscoelastic flows in confined geometries. 

\section{Acknowledgment}

We kindly acknowledge funding from the European Research Council under the Europeans Community's Seventh Framework Programme (FP7/2007-2013)/ERC Grant Agreement No. 279004; We acknowledge computational support from CINECA and from PRACE-7th Call Grant MULTIPORE. We also acknowledge L. Biferale and A. Scagliarini for useful discussions.

\appendix

\section{Hybrid Lattice Boltzmann Models (LBM) - Finite Difference Scheme for dilute Polymeric solutions}\label{appendix}

In this appendix we report the essential technical details of the numerical scheme used. We refer the interested reader to a dedicated paper~\cite{SbragagliaGupta} where all the technical details are reported and the model benchmarked by characterizing the rheological behaviour of dilute homogeneous solutions in various configurations, including steady shear flow, elongational flows, transient shear and oscillatory flows. The LBM equations evolve in time the discretized probability density function $f_{\zeta \ell}({\bm{x}},t)$ to find at position ${\bm{x}}$ and time $t$ a fluid particle of component $\zeta=A,B$ (the two components indicate the droplet (D) or the matrix (M) Newtonian phases in Eqs.~(\ref{EQ}) and (\ref{EQB})) with velocity  ${\bm{c}}_{\ell}$ according to the updating scheme
\begin{equation}\label{EQ:LBapp}
f_{\zeta \ell} ({\bm{x}} + {\bm{c}}_{\ell} , t + 1)-f_{\zeta \ell} ({\bm{x}}, t) = \sum_{j} {\cal L}_{\ell j}(f_{\zeta j}-f^{(eq)}_{\zeta j}) + \Delta^{g}_{\zeta \ell}
\end{equation}
where the lattice time step $\Delta t$ has been set to a unitary value for simplicity. The (linear) collisional operator in the rhs of Eq.~\eqref{EQ:LBapp} expresses the relaxation of the probability distribution function towards the local equilibrium $f^{(eq)}_{\zeta \ell}$. The expression for the equilibrium distribution is a result of the projection onto the lower order Hermite polynomials~\cite{Dunweg07,DHumieres02} and the weights $w_{\ell}$ are {\it a priori} known through the choice of the quadrature
\be\label{feq}
f_{\zeta \ell}^{(eq)}=w_{\ell} \rho_{\zeta} \left[1+\frac{{\bm{u}} \cdot {\bm{c}}_{\ell}}{c_s^2}+\frac{{\bm{u}}{\bm{u}}:({\bm{c}}_{\ell}{\bm{c}}_{\ell}-{\Id})}{2 c_s^4} \right]
\ee
\begin{equation}\label{weights}
w_{\ell}=
\begin{cases}
1/3 & \ell=0\\
1/18 & \ell=1\ldots6\\
1/36 & \ell=7\ldots18
\end{cases},
\end{equation}
where $c_s$ is the isothermal speed of sound (a constant in the model) and ${\bm{u}}$ is the fluid velocity. Our implementation features a D3Q19 model with 19 velocities
\begin{equation}\label{velo}
{\bm{c}}_{\ell}=
\begin{cases}
(0,0,0) & \ell=0\\
(\pm 1,0,0), (0,\pm 1,0), (0,0,\pm 1) & \ell=1\ldots6\\
(\pm 1,\pm 1,0), (\pm 1,0,\pm 1), (0,\pm 1,\pm 1)  & \ell=7\ldots18
\end{cases}.
\end{equation}
The operator ${\cal L}_{\ell j}$ in Eq.~\eqref{EQ:LBapp} is the same for both components and is constructed to have a diagonal representation in the so-called {\it mode space}:  the basis vectors ${\bm{e}}_{k}$ ($k=0,...,18$) of mode space are constructed by orthogonalizing polynomials of the dimensionless velocity vectors~\cite{Dunweg07,DHumieres02}. The basis vectors are used to calculate a complete set of moments, the so-called modes $m_{\zeta k}=\sum_{\ell} {\bm{e}}_{k \ell} f_{\zeta \ell}$ ($k=0,...,18$). The lowest order modes are associated with the hydrodynamic variables. In particular, the zero-th order moment gives the densities for both components, $\rho_{\zeta}=m_{\zeta 0}=\sum_{\ell} f_{\zeta \ell}$, with the total density given by $\rho=\sum_{\zeta}m_{\zeta 0} =\sum_{\zeta}\rho_{\zeta}$. The next three moments $\tilde{\bm{m}}_{\zeta}=(m_{\zeta 1}, m_{\zeta 2}, m_{\zeta 3})$, when properly summed over all the components, are related to the velocity of the mixture 
\be\label{totmom}
{\bm{u}} \equiv \frac{1}{\rho}\sum_{\zeta} \tilde{\bm{m}}_{\zeta}   +\frac{\bm{g}}{2 \rho} = \frac{1}{\rho}\sum_{\zeta} \sum_{\ell} f_{\zeta i} {\bm{c}}_{\ell}+\frac{\bm{g}}{2 \rho}.
\ee
The other modes are the bulk and the shear modes (associated with the viscous stress tensor), and four groups of kinetic modes which do not emerge at the hydrodynamic level~\cite{Dunweg07,DHumieres02}. Since the operator ${\cal L}_{\ell j}$ is  diagonal in mode space, the collisional term describes a linear relaxation of the non-equilibrium modes
\be\label{MODES}
m^{*}_{\zeta k}=(1+\lambda_k)m_{\zeta k}+m_{\zeta k}^{g}
\ee
where the $*$ indicates the post-collisional mode and where the relaxation frequencies $-\lambda_k$ (i.e. the eigenvalues of $-{\cal L}_{\ell j}$) are related to the transport coefficients of the modes. The term $m_{\zeta k}^{g}$ is related to the $k$-th moment of the forcing source $\Delta_{\zeta \ell}^{g}$ associated with a forcing term with density ${\bm{g}}_{\zeta}$. The term ${\bm{g}}=\sum_{\zeta} {\bm g}_{\zeta}$ in Eq.~\eqref{totmom} refers to all the contributions coming from internal and external forces. While the forces have no effect on the mass density, they transfer an amount ${\bm{g}}_{\zeta}$ of total momentum to the fluid in one time step. The forcing term is determined in such a way that the hydrodynamic Eqs.~(\ref{eq:2})-(\ref{eq:3}) are recovered, and can be written as~\cite{Guo}
\begin{equation}
\Delta_{\zeta \ell}^{g}=\frac{w_{\ell}}{c_s^2} \left(\frac{2+\lambda_M}{2}\right) {\bm{g}}_{\zeta} \cdot {\bm{c}}_{\ell} +\frac{w_{\ell}}{c_s^2} \left[\frac{1}{2c_s^2} {\bm{G}} : ({\bm{c}}_{\ell} {\bm{c}}_{\ell}-c_s^2 {\Id} ) \right],
\end{equation}
where the tensor ${\bm G}$ is defined as
\begin{equation}
{\bm G}=\frac{2+\lambda_s}{2}\left({\bm u} {\bm g} + ({\bm u}  {\bm g})^T-\frac{2}{3} {\Id} ({\bm u} \cdot {\bm g}) \right)+\frac{2+\lambda_b}{3} {\Id} ({\bm u} \cdot {\bm g}).
\end{equation}
In the above equations we have used explicitly the relaxation frequencies of the momentum ($-\lambda_M$), bulk ($-\lambda_b$) and shear ($-\lambda_s$) modes.  Using the LBM we are able to reproduce the continuity equations and the Navier Stokes equations for the total momentum
\be\label{eq:2}
\partial_t \rho_{\zeta}+ {\bm \nabla} \cdot (\rho_{\zeta} {\bm u}) = {\bm \nabla} \cdot {\bm D}_{\zeta},
\ee
\be\label{eq:3}
\rho \left[ \partial_t \bm u + ({\bm u} \cdot {\bm \nabla}) \bm u \right]= -{\bm \nabla}p+ {\bm \nabla} \left[ \eta_{s} \left( {\bm \nabla} {\bm u}+({\bm \nabla} {\bm u})^{T}-\frac{2}{3} {\Id} ({\bm \nabla} \cdot {\bm u}) \right) +\eta_{b}{\Id} ({\bm \nabla} \cdot {\bm u}) \right] + {\bm g}
\ee
where $\eta_s$, $\eta_b$ are the shear and bulk viscosities, respectively. In Eq.~\eqref{eq:3}, $p=\sum_{\zeta} p_{\zeta}=\sum_{\zeta} c_s^2 \rho_{\zeta}$ is the internal (ideal) pressure of the mixture. The quantity ${\bm D}_{\zeta}$ represents the diffusion flux of one component into the other
\begin{equation}\label{eq:comp_Pi}
{\bm D}_{\zeta}=\mu \left[\left({\bm \nabla} p_{\zeta}-\frac{\rho_{\zeta}}{\rho}{\bm \nabla} p\right)-\left({\bm g}_{\zeta}-\frac{\rho_{\zeta}}{\rho} {\bm g} \right) \right]
\end{equation}
with $\mu$ a mobility parameter regulating the intensity of such diffusion flux. As for the internal forces, we will use the ``Shan-Chen'' model~\cite{SC93} for multicomponent mixtures
\begin{equation}\label{eq:SCforce}
{\bm g}_{\zeta}({\bm{x}}) =  - {\cal G} \rho_{\zeta}({\bm{x}}) \sum_{\ell} \sum_{\zeta'\neq \zeta} w_{\ell} \rho_{\zeta^{\prime}} ({\bm{x}}+\bm{c}_{\ell}) {\bm c}_{\ell} \hspace{.2in} \zeta,\zeta^{\prime}=A,B
\end{equation}
where ${\cal G}$ is a parameter that regulates the interactions between the two components. The sum in Eq.~\eqref{eq:SCforce} extends over a set of interaction links coinciding with those of the LBM dynamics (see Eq.~\eqref{velo}). When the coupling strength parameter ${\cal G}$ is sufficiently large, demixing occurs and the model can describe stable interfaces with a surface tension. The resulting physical domain is partitioned into two different phases, each with a majority of one of the two components, with the interface between the two phases described as a thin layer where the fluid properties change smoothly. The effect of the internal forces can be recast into the gradient of the pressure tensor ${\bm P}^{(\mbox{\tiny{int}})}$~\cite{SbragagliaBelardinelli}, thus modifying the internal pressure of the model, i.e. ${\bm P} = p \, {\Id}+{\bm P}^{(\mbox{\tiny{int}})}$, with 
\be\label{PT}
{\bm P}^{(\mbox{\tiny{int}})}({\bm x}) =\frac{1}{2} {\cal G} \rho_{A}({\bm x})\sum_{\ell} w_{\ell} \rho_{B}({\bm x}+{\bm c}_{\ell}) {\bm c}_{\ell} {\bm c}_{\ell}+\frac{1}{2} {\cal G} \rho_{B}({\bm x})\sum_{\ell} w_{\ell} \rho_{A}({\bm x}+{\bm c}_{\ell}){\bm c}_{\ell} {\bm c}_{\ell} 
\ee
Upon Taylor expanding the expression (\ref{PT}), we get a bulk pressure contribution $P=p+c_s^2{\cal G}\rho_A \rho_B$ (which is the bulk pressure appearing in Eqs.~\eqref{EQ} and \eqref{EQB}) and a contribution proportional to the density gradients, which are responsible for the surface tension at the non ideal interface. A proper tuning of the density gradients in contact with the wall allows for the modelling of the wetting properties. In all the simulations described in this paper, the resulting contact angle for a droplet placed in contact with the solid walls is $\theta_{\mbox{\tiny{wet}}}=90^{\circ}$ (i.e. neutral wetting).  The relaxation frequencies of the momentum, bulk and shear modes in (\ref{EQ:LBapp}) are related to the transport coefficients of hydrodynamics as
\begin{equation}\label{TRANSPORTCOEFF}
\mu=-\left(\frac{1}{\lambda_M}+\frac{1}{2} \right) \hspace{.2in} \eta_s=-\rho c_s^2 \left(\frac{1}{\lambda_s}+\frac{1}{2} \right) \hspace{.2in} \eta_b=-\frac{2}{3}\rho c_s^2  \left(\frac{1}{\lambda_b}+\frac{1}{2} \right).
\end{equation}
For the numerical simulations presented we have used ${\cal G}=1.5$ lbu in (\ref{eq:SCforce}) corresponding to a surface tension $\sigma=0.1$ lbu and associated bulk densities $\rho_A=2.0$ lbu and $\rho_B=0.1$ lbu in the $A$-rich region (see Fig.~\ref{fig:1}). The relaxation frequencies in (\ref{TRANSPORTCOEFF}) are such that $\lambda_M=-1.0$ lbu and $\lambda_s=\lambda_b$, which reproduces the viscous stress tensor given in Eqs.~(\ref{EQ}) and (\ref{EQB}).  The viscous ratio of the LBM fluid is changed by letting $\lambda_s$ depend on space 
\be
-\rho c_s^2 \left(\frac{1}{\lambda_s}+\frac{1}{2} \right)=\eta_s=\eta_A (f_+(\phi))+\eta_B (f_{-}(\phi))
\ee
where $\phi=\phi({\bm x})=\frac{(\rho_A({\bm x})-\rho_B({\bm x}))}{(\rho_A({\bm x})+\rho_B({\bm x}))}$ represents the order parameter. The functions $f_{\pm}(\phi)$ are chosen as
\be
f_{\pm}(\phi)=\left(\frac{1 \pm \tanh(\phi/\xi)}{2}\right)
\ee
which allows to recover, in the two bulk phases, the Newtonian part of the Navier Stokes equations reported in Eqs.~(\ref{NS}) and (\ref{EQB}) with shear viscosities $\eta_A$ and $\eta_B$. The smoothing parameter $\xi$ is chosen sufficiently small so as to recover a good matching with the analytical prediction of the droplet deformation (See Fig.~\ref{fig:2}).\\
As for the polymer evolution given in Eqs.~\eqref{FENE} and \eqref{FENEb}, we are following the two References~\cite{perlekar06,vaithianathan2003numerical} to solve the FENE-P equation. We maintain the symmetric-positive-definite nature of conformation tensor at all times by using the Cholesky-decomposition scheme~\cite{perlekar06,vaithianathan2003numerical}. The polymer stress $f(r_P){\bm {\mathcal C}}$ is computed from the FENE-P evolution equation and used to change the shear modes of the LBM~\cite{SbragagliaGupta,Dunweg07,DHumieres02}. In the spirit of the diffuse interface models proposed by Yue {\it et al.}~\cite{Yue04}, the feedback of the polymers is modulated in space with the function $f_{+}(\phi)$ 
\be
\rho \left[ \partial_t \bm u + ({\bm u} \cdot {\bm \nabla}) \bm u \right]= -{\bm \nabla} {\bm P}+ {\bm \nabla} \left[(\eta_A f_+(\phi)+\eta_B f_{-}(\phi)) ({\bm \nabla} {\bm u}+({\bm \nabla} {\bm u})^{T} ) \right]+\frac{\eta_P}{\tau_P}{\bm \nabla} [f(r_P){\bm {\mathcal C}} f_{+}(\phi) ]
\ee
which recovers Eq.~(\ref{EQ}) in the droplet phase with a Newtonian matrix phase. Consistently, if the polymer feedback stress is modulated in space with the function $f_{-}(\phi)$, we recover a case with matrix viscoelasticity and a Newtonian droplet (see Sec.~\ref{sec:dropletdeformation}).
\section*{References}

\bibliography{archive}

\end{document}